\documentclass[a4paper,12pt]{article}
\usepackage{newlfont}
\usepackage{amssymb}
\usepackage{amsfonts}
\usepackage{amsmath}

\newtheorem{theo}{Theorem}
\newtheorem{defi}{Definition}
\newtheorem{Collorary}{Collorary}
\newtheorem{ex}{Example}
\newtheorem{fact}{Fact}
\newcommand{\tr}{\mbox{Tr}}
\newcommand{\bra}[1]{\mbox{$\langle #1 |$}}
\newcommand{\ket}[1]{\mbox{$| #1 \rangle$}}
\newcommand{\bk}[2]{\ensuremath{\langle #1 | #2 \rangle}}
\newcommand{\kb}[2]{\ensuremath{| #1 \rangle\!\langle #2 |}}

\begin{document}

\title{Symplectic geometry of entanglement}
\author{Adam Sawicki$^{1}$, Alan Huckleberry$^{2}$, and Marek Ku\'s$^{1}$
\\ \\
$^1$Center for Theoretical Physics, Polish Academy of Sciences\\ Al.
Lotnik\'ow 32/46,
02-668 Warszawa, Poland  \\ \\
$^2$Fakult\"at f\"ur Mathematik, Ruhr-Universit\"at Bochum,\\
D-44780 Bochum, Germany \\ \\
}
\maketitle

\begin{abstract}

We present a description of entanglement in composite quantum systems in
terms of symplectic geometry. We provide a symplectic characterization of
sets of equally entangled states as orbits of group actions in the space of
states. In particular, using Kostant-Sternberg theorem, we show that
separable states form a unique K\"ahler orbit, whereas orbits of entanglement
states are characterized by different degrees of degeneracy of the canonical
symplectic form on the complex projective space.  The degree of degeneracy
may be thus used as a new geometric measure of entanglement and we show how
to calculate it for various multiparticle systems providing also simple
criteria of separability. The presented method is general and can be applied
also under different additional symmetry conditions stemming, eg. from the
indistinguishability of particles.
\end{abstract}

\section{Introduction}

Quantum entanglement - a direct consequence of linearity of quantum mechanics
and the superposition principle - is one of the most intriguing phenomena
distinguishing quantum and classical description of physical systems. Quantum
states which are entangled posses features unknown in the classical world,
like the seemingly paradoxical non-local properties exhibited by the famous
Einstein-Podolsky-Rosen analysis of completeness of the quantum theory.
Recently, with the development of quantum information theory they came to
prominence as the main resource for several applications aiming at speeding
up and making more secure information transfers (see e.g.\
\cite{horodecki09}).

Pure states which are not entangled are called separable and for
systems of $N$ distinguishable particles they are, by definition,
described by simple tensors in the Hilbert space of the whole
system,
$\mathcal{H}=\mathcal{H}_1\otimes\cdots\otimes\mathcal{H}_N$, where
$\mathcal{H}_k$ are the single-particle spaces. For
indistinguishable particles such a definition lacks sense -
indistinguishability enforces symmetrization or antisymmetrization
of the state vectors. In effect nearly all states are not simple
tensors, in fact the relevant Hilbert spaces of such systems are not
longer tensor products, but rather their symmetric or antisymmetric
subspaces. In these cases one modifies the original definition of
separability and adapts it according to symmetry (see below).

The concept of separability (or equivalently nonentanglement) can be
in a natural way extended to mixed states by first identifying pure
states with projections on their directions (i.e.\ rank-one
orthogonal projections) and then defining mixed separable states as
convex combinations of pure separable ones. Mixed states which are
not separable are, consequently, called entangled.

Separability of a state remains unaffected under particular class of
transformations allowed by quantum mechanics. Thus, for example, a separable
state of distinguishable particles remains separable when we act on it by a
unitary operator $U=U_1\otimes\cdots\otimes U_N$ where $U_k$ are unitaries
acting in the single-particle spaces. One can find appropriate classes of
unitary operators preserving separability also in the cases of
indistinguishable particles. Going one step further one may analyze how
actions of separability-preserving unitaries stratifies into their orbits the
whole space of states (pure or mixed) of a composite quantum system. To treat
all the cases in a unified way we may consider a general situation in which a
compact group $K$ acts on some manifold $M$. The manifold in question will
then depend on the considered system. For pure states it will be the
projectivisation $\mathbb{P}(\mathcal{H})$ of the Hilbert space $\mathcal{H}$
in the case of distinguishable particles or the projectivisation of an
appropriate symmetrization (for bosons) or antisymmetrization (for fermions)
of $\mathcal{H}$. In all cases the manifold $M$ is naturally equipped with
some additional structure. In our investigations it will be a symplectic
structure inherited from the natural one existing on every complex Hilbert
space. Orbits of $K$ being submanifolds of $M$ might also, under special
circumstances, inherit the symplectic structure or in addition respect the
underlying complex structure of $\mathcal{H}$ and become K\"ahlerian. Form
this point of view we want to consider several problems.
\begin{enumerate}
\item How symplectic and non-symplectic orbits of the $K$ action on
$\mathbb{P}(V)$ stratify the set of pure states?
\item What is the meaning (for the entanglement properties) of the fact
    that the orbit through a particular pure state is or is not
    symplectic?
\end{enumerate}

In the next section we start with relevant definitions of separability and
entanglement for distinguishable as well as indistinguishable particles. When
giving definitions we concentrate on $N=2$, i.e.\ on two-partite systems, but
the general reasonings for larger $N$ remains very similar. To make the paper
reasonably self-contained we devote a few further sections and the Appendix
to a presentation of some tools from the Lie-group representation theory and
the symplectic geometry most important in our investigations.

\section{Separable and entangled states}
\label{sec:sep}

Let $\mathcal{H}$ be an $N$-dimensional Hilbert space. By choosing
an orthonormal basis in $\mathcal{H}$ we will identify it with
$\mathbb{C}^N$ equipped with the standard Hermitian product.

A state is a positive, trace-one linear operator on $\mathcal{H}$,
\begin{equation}\label{state}
\rho:\mathcal{H}\rightarrow\mathcal{H},\quad
\forall_{x\in\mathcal{H}}\bra{x}\rho\ket{x}\ge 0, \quad \tr\rho=1.
\end{equation}
We use the standard Dirac notation: $\ket{x}$ is an element of
$\mathcal{H}$, and $\bra{x}$ - the element of the dual space
$\mathcal{H}^*$ corresponding to $\ket{x}$ \textit{via} the scalar
product $\bk{\,\cdot\,}{\,\cdot}$ on $\mathcal{H}$. A state is, by
definition, \textit{pure} if it is a rank-one projection,
\begin{equation}\label{pure}
\rho=\rho^2,
\end{equation}
otherwise it is called \textit{mixed}. A pure state can be thus
written in the form $\rho=\kb{x}{x}/\bk{x}{x}:=P_x$ for some
$x\in\mathcal{H}$, hence it can be identified with a point in the
projective space $\mathbb{P}(\mathcal{H})$.

\subsection{Separable and entangled states of two distinguishable particles}

The Hilbert space for a composite system of two distinguishable
particles is the tensor product of the Hilbert spaces of the
subsytems,
\begin{equation}\label{Hcomp}
\mathcal{H}=\mathcal{H}_1\otimes\mathcal{H}_2, \quad \mathcal{H}_1
\simeq\mathbb{C}^N,\quad \mathcal{H}_2\simeq\mathbb{C}^M.
\end{equation}
A pure state $\rho$ is called \textit{separable} or, equivalently,
\textit{nonentangled} if and only if it is a tensor product of pure states of
the subsystems,
\begin{equation}\label{pseparable}
\rho=P_x\otimes P_y, \quad
\ket{x}\in\mathcal{H}_1,\,\,\ket{y}\in\mathcal{H}_2,
\end{equation}
otherwise it is called \textit{entangled}. A mixed state is, by
definition, separable if it is a convex combination of pure
separable states \cite{werner89},
\begin{equation}\label{mseparable}
\rho=\sum_i p_i P_{x_i}\otimes P_{y_i}, \quad
\ket{x_i}\in\mathcal{H}_1,\quad \ket{y_i}\in\mathcal{H}_2, \quad
p_i>0,\quad \sum_i p_i=1.
\end{equation}

From the physical point of view it is often desirable to define \textit{how
strongly entangled} is a particular state $\rho$. Although such a
quantification of entanglement is not universal, especially for systems with
more then two constituents and can be constructed on the basis of different
(measured in an actual experiment) properties of entangled states, it should
always ascribe the same \textit{amount of entanglement} to states differeing
by \textit{local quantum operations}, i.e.\ by a conjugation by direct
product of the unitary groups $U(\mathcal{H}_1)\times U(\mathcal{H}_2)$,
\begin{equation}\label{loc}
\rho\mapsto U_1\otimes U_2\rho\, U_1^\dagger\otimes U_2^\dagger.
\end{equation}

\subsection{Separable and entangled states of two indistinguishable
particles}\label{subsec:ind-def}

For indistinguishable particles the Hilbert space of a composite,
two-partite system is no longer the tensor product of the Hilbert
spaces of the subsystems but,
\begin{enumerate}
\item the antisymmetric part of the tensor product in the case of
fermions,
\begin{equation}\label{Hfermion}
\mathcal{H}_F=\bigwedge{}^2\left(\mathcal{H}_1\right),
\end{equation}
\item the symmetric part of the tensor product in the case of bosons,
\begin{equation}\label{Hboson}
\mathcal{H}_B=\mathrm{Sym}^2\left(\mathcal{H}_1\right),
\end{equation}
\end{enumerate}
where $\mathcal{H}_1\simeq\mathbb{C}^M$ is the, so called,
\textit{one-particle Hilbert space}, i.e.\ the Hilbert space of a
single particle.

In the fermionic state there is a natural way of defining pure nonentangled
states: a state $\rho$ is \textit{nonentangled} if and only if it is an
orthogonal projection on an antisymmetric part of the tensor product of two
vectors from $\mathcal{H}_1$ \cite{sckll01,eckert02}. Otherwise $\rho$ is
called \textit{entangled}. This definition, which can be in an obvious way
extended to multipartite systems, is equivalent to the one proposed in
\cite{sckll01} and \cite{eckert02}.

Interestingly, a completely analogous definition for bosons, identifying
nonentangled pure states with orthogonal projections of simple tensors on the
symmetric part of the tensor product of two (ore more when the number of
subsystems exceeds two) copies of $\mathcal{H}_1$, leads to some unexpected
consequences: there are two geometrically inequivalent types of nonentangled
bosonic states. We will return to the problem in Section~\ref{sec:bosons}.
There exists an alternative solution which is tantamount to defining as
nonentangled only those states which are products of two (or more) copies of
the same state from $\mathcal{H}_1$. Both definitions, supported by physical
arguments, were employed in the literature of the subject. In
\cite{ghirardi02} (see also \cite{li01}) a concept of `complete system of
properties' of a subsystem was used to introduce a definition of
nonentanglement of the first of the above described kinds, whereas in
\cite{paskauskas01} it was pointed that the second kind of definition assures
that nonentangled states can not be used to perform such clearly
`non-classical' task like e.g.\ teleportation, which definitely remains in
accordance with the basic intuition connecting no-entanglement with the
classical world. The second definition of nonentangled bosonic states was
also proposed in \cite{eckert02}, based on slightly different arguments.

Mixed nonentangled states for fermions and bosons are defined, as in the case
of distinguishable particles, as convex combinations of pure nonentangled
states.

As in the case of distinguishable particles the physically
interesting \textit{amount of entanglement} is invariant under the
action of $U(\mathcal{H}_1)$ acting in the one-particle space
$\mathcal{H}_1$.

\section{Pure nonentangled states as coherent states}
\label{sec:cohstates}

In all three cases of distinguishable particles, fermions, and bosons, the
pure nonentangled states, treated as points in appropriate projective spaces,
form a set invariant under the action of an appropriate compact, semisimple
group $K$ irreducibly represented on some Hilbert space $\mathcal{H}$
\cite{klyachko08,bengtsson07,kb09}. This observation is in accordance with an
intuition that entanglement properties of a state should not change under
`local' transformations allowed by quantum mechanics and symmetries of a
system. Thus for example, for two distinguishable particles in two distant
laboratories, local transformations can consist of independent quantum
evolutions of each particle. This paradigm does not apply to
indistinguishable particles when, in order to keep the exchange symmetry
untouched, both particles must undergo the same evolution. Thus,
\begin{enumerate}
\item For distinguishable particles,
\begin{equation}\label{dcoh}
K=SU(N)\times SU(M), \quad \mathcal{H}=\mathbb{C}^N\otimes\mathbb{C}^M.
\end{equation}
\item For fermions,
\begin{equation}\label{fcoh}
K=SU(N), \quad \mathcal{H}=\bigwedge{}^2\left(\mathbb{C}^N\right),
\end{equation}
\item For bosons,
\begin{equation}\label{fbos}
K=SU(N),\quad \mathcal{H}=\mathrm{Sym}^2\left(\mathbb{C}^N\right).
\end{equation}
\end{enumerate}
In all cases the nonentangled pure states are distinguished as forming some
unique orbit of the underlying group action \cite{kostant82,guillemin84}. The
orbit in question appears in the literature in several contexts and customary
its points are called coherent states, or the coherent states `closest to
classical states' \cite{perelomov86}) A precise characteristic of the orbit,
as well as its distinguished features from the view of entanglement theory
will be discussed below.

\section{A short review of the representation theory}
Let us remind some fundamentals of the representation theory for semisimple
Lie groups and algebras useful in next sections \cite{hall03}.

In the following we denote by $K$ a simply connected compact Lie
group and by $\mathfrak{k}$ its Lie algebra. It is standard fact
that representations of $K$ are in one to one correspondence with
representations of $\mathfrak{k}$. They both posses complete
reducibility property, i.e., decompose as direct sums of irreducible
ones, and can be made unitary by an appropriate choice of the scalar
product in the carrier space. Let $\mathfrak{k}^{\mathbb{C}}$ be the
complexification of $\mathfrak{k}$. It is also well known that
irreducible representations of $\mathfrak{k}$ and
$\mathfrak{k}^{\mathbb{C}}$ are in one to one correspondence and
that $\mathfrak{k}^{\mathbb{C}}$ is a semisimple complex Lie
algebra.
\begin{ex} Consider $K=SU(n)$ which is simply connected and compact.
Then $\mathfrak{k}=\mathfrak{su}(n)$ and
$\mathfrak{k}^{\mathbb{C}}=\mathfrak{sl}(n,\mathbb{C})$.
\end{ex}

\subsection{Adjoint representation of $\mathfrak{k}^{\mathbb{C}}$}
The adjoint representation of $\mathfrak{k}^{\mathbb{C}}$ is defined
as
\begin{eqnarray}
\mathrm{ad}:\mathfrak{k}^{\mathbb{C}}\rightarrow
\mathfrak{gl}(\mathfrak{k}^{\mathbb{C}}), \label{defad1}\\
\mathrm{ad}_X(Y)=[X,Y]. \label{defad2}
\end{eqnarray}
This representation plays a key role in understanding all other
representations of $\mathfrak{k}^{\mathbb{C}}$. Let us fix a maximal
commutative subalgebra $\mathfrak{t}$ of $\mathfrak{k}$ then
$\mathfrak{h}=\mathfrak{t}^{\mathbb{C}}=\mathfrak{t}+i\mathfrak{t}$ is a
Cartan subalgebra of $\mathfrak{k}^{\mathbb{C}}$. Since $\mathfrak{h}$ is the
maximal commutative subalgebra of $\mathfrak{k}^{\mathbb{C}}$ with the
property that for every $H\in\mathfrak{h}$ the operator $\mathrm{ad}_H$ is
diagonalizable (this is a consequence of the assumed semisimplicity of $K$),
we can decompose $\mathfrak{k}^{\mathbb{C}}$ as a direct sum of root spaces
with respect to $\mathfrak{h}$,
\begin{eqnarray}\label{decomp1}
\mathfrak{k}^{\mathbb{C}} =\mathfrak{h}\oplus\bigoplus_\alpha
\mathfrak{g}_\alpha ,
\end{eqnarray}
where $\alpha:\mathfrak{h}\rightarrow\mathbb{C}$ range over linear
functionals (called roots) for which there exist
$X\in\mathfrak{k}^{\mathbb{C}}$ such that
\begin{eqnarray}\label{droots}
\mathrm{ad}_H(X)=\alpha(H)X \quad \forall H\in\mathfrak{h}
\end{eqnarray}
Space $\mathfrak{g}_\alpha$ consists of the elements $X$ with the
above property. It is a standard fact that if $\alpha$ is a root
then $-\alpha$ is also a root and that
$[\mathfrak{g}_\alpha,\mathfrak{g}_\beta]=0$ or
$[\mathfrak{g}_\alpha,\mathfrak{g}_\beta]=\mathfrak{g}_{\alpha+\beta}$.
Moreover all $\mathfrak{g}_\alpha$ are one dimensional. We may
introduce the notion of a positive root by first choosing an
arbitrary basis consisting of roots in the space spanned by them,
and then defining positive roots as those with only positive
coefficients in the decomposition in the chosen basis. The weight
space decomposition of $\mathfrak{k}^{\mathbb{C}}$ can be then
written as
\begin{eqnarray}\label{decomp2}
\mathfrak{k}^{\mathbb{C}} =\mathfrak{n}_-\oplus\mathfrak{h}\oplus
\mathfrak{n}_+
\end{eqnarray}
where the direct sums of the negative and positive root spaces,
$\mathfrak{n}_-$ and $\mathfrak{n}_+$ are nilpotent Lie algebras. In
the defining representation of $\mathfrak{sl}(n,\mathbb{C})$ as
$N\times N$ complex traceless matrices, the most natural choice of
positive roots is that which leads to $\mathfrak{n}_-$ and
$\mathfrak{n}_+$ as, respectively, lower and upper triangular
matrices.

It is a key fact that we can choose bases $E_\alpha$ of the root
spaces $\mathfrak{g}_\alpha$ and define
$H_\alpha=[E_{-\alpha},E_\alpha]$ so that
$\{E_{-\alpha},H_\alpha,E_\alpha\}$ is the standard basis for
$\mathfrak{sl}_2(\mathbb{C})$. We will denote it by
$\mathfrak{sl}_2(\alpha)$ and by $\mathfrak{su}_2(\alpha)$ the
corresponding $\mathfrak{su}_2$-triple
$\{E_{-\alpha}-E_\alpha,iH_\alpha,i(E_{-\alpha}+E_\alpha)\}$.

\subsection{General case}\label{generalcase}
It is enough to restrict our attention to irreducible
representations as $K$ is a compact and simply connected Lie group.
Given any representation of $\mathfrak{h}$ on complex vector space
$V$ one decomposes $V$ as a direct sum:
\begin{eqnarray}\label{decomp2a}
V=\oplus V_\lambda
\end{eqnarray}
where isotypical components $V_\lambda$ are weight spaces. In other
words:
\begin{eqnarray}\label{defweight}
\xi.v=\lambda(\xi)v\quad \forall
\xi\in\mathfrak{h}\quad\mathrm{and}\quad v\in V_\lambda,
\end{eqnarray}
where the linear functionals $\lambda$ are called weights and
vectors $v$ - the corresponding weight vectors. Every irreducible
representation of $\mathfrak{k}^{\mathbb{C}}$ is the so called
highest weight cyclic representation. The most important facts we
will use in next sections are:
\begin{itemize}
\item $E_\alpha.V_\lambda\subset V_{\lambda+\alpha}$
\item $[E_{\alpha},E_\beta].V_\lambda\subset V_{\lambda+\alpha+\beta}$
\end{itemize}
where $E_\alpha\in\mathfrak{g}_\alpha$ and
$E_\beta\in\mathfrak{g}_\beta$.

\section{Symplectic orbits of group actions}
\label{sec:symp}

In the following we will need a couple of facts about actions of Lie
groups on symplectic manifolds (see e.g. \cite{guillemin84}).

Let us denote by $(M,\omega)$ a symplectic manifold, i.e.\ $M$ is a
manifold and $\omega$ is a nondegenerate, closed ($d\omega=0$)
two-form.

Let a compact semisimple group $K$ act on $M$ \textit{via}
syplectomorhisms, $K\times M\ni(g,x)\mapsto\Phi_g(x)\in M$,
$\Phi_g^\ast\omega=\omega$. We denote by $\mathfrak{k}^\ast$ the
space dual to $\mathfrak{k}=Lie(K)$.

Let $\xi\in\mathfrak{k}$. We define a vector field $\hat{\xi}$
\begin{equation}\label{fvf}
\hat\xi(x)=\frac{d}{dt}\bigg|_{t=0} \Phi_{\exp t\xi}(x).
\end{equation}
Since the action of the group is Hamiltonian (which is true for a
semisimple $K$), for each $\xi\in\mathfrak{k}$ there exists a
Hamilton function $\mu_\xi:M\rightarrow \mathbb{R}$ for $\hat{\xi}$,
i.e.
\begin{equation}\label{moment1}
d\mu_\xi=\imath_{\hat{\xi}}\,\omega:=\omega(\xi,\cdot).
\end{equation}
The function can be chosen to be linear in $\xi$, i.e.
\begin{equation}\label{moment-lin}
\mu_\xi(x)=\langle\mu(x),\xi\rangle,\quad
\mu(x)\in\mathfrak{k}^\ast,
\end{equation}
where $\langle\,,\rangle$ is the pairing between $\mathfrak{k}$ and
its dual $\mathfrak{k}^\ast$. The map $\mu_\xi$ defines thus by
(\ref{moment-lin}) a map $\mu:M\rightarrow\mathfrak{k}^\ast$. We can
chose $\mu$ to be equivariant with respect to the coadjoint action
of $K$ \cite{kirillov04}, i.e.\
\begin{equation}\label{moment-equiv}
\mu\left(\Phi_g(x)\right)=\mathrm{Ad}^\ast_g \mu(x),
\end{equation}
where the coadjoint action $\mathrm{Ad}^\ast_g$ on
$\mathfrak{k}^\ast$ is defined \textit{via}
\begin{equation}\label{Adast}
\langle\mathrm{Ad}^\ast_g\alpha,h\rangle=\langle\alpha,\mathrm{Ad}_{g^{-1}}
h\rangle=\langle\alpha,g^{-1}hg\rangle, \quad g\in K, \quad
h\in\mathfrak{k}, \quad \alpha\in\mathfrak{k}^\ast,
\end{equation}
and $\mathrm{Ad}$ is the adjoint representation of $K$,
\begin{eqnarray}\label{Ad}
\mathrm{Ad}:K\rightarrow Gl(\mathfrak{k}) \nonumber \\
\mathrm{Ad}(g)X=gXg^{-1}:=\frac{d}{dt}\bigg|_{t=0}g\,\exp tX\, g^{-1} .
\end{eqnarray}

The above constructed $\mu$ is called the \textit{momentum map}.

The goal is now to describe the criterion for $K$-orbit to be symplectic. Let
$N=K.x$ be the orbit through a point $x\in M$. Denote by $\omega_N$ the
restriction of the symplectic form $\omega$ to $N$. This form may, and in
fact usually does, have a certain degree of degeneracy. Denote by $D_x$ the
subspace of tangent vectors which are $\omega_N$-orthogonal to the full space
$T_xN$. Since the $K$ action is symplectic we have
$\Phi_{g\ast}(D_x)=D_{\Phi_g(x)}$ which means that degree of degeneracy is
constant on the orbit $N$. This fact will turn out to be very important in
the context of entanglement measure. Now because of (\ref{moment-equiv}) $\mu
(N)=\mathcal{O}$ is a coadjoint orbit in $\mathfrak{k}^\ast$ and thus is
symplectic with respect to the canonical form $\omega_{\mathcal{O}}$ (see
Appendix). We also have $(\mu|_N)^\ast(\omega_{\mathcal{O}})=\omega_N$ which
means that the tangent spaces of the fibers of $\mu|_N$ are exactly the
degeneracy spaces $N_x$. Indeed, if $u\in D_x$ then for an arbitray $v\in T_x
N$
\begin{equation}\label{why}
0=\omega_N(u,v)=\omega_{\mathcal{O}}((\mu|_{N})_{\ast}u,(\mu|_{N})_{\ast}v).
\end{equation}
But $\omega_{\mathcal{O}}$ is nondegenerate and
$T_{\mu(x)}\mathcal{O}=(\mu|_{N})_{\ast}T_xN$. Thus
$(\mu|_{N})_{\ast}u=0$ whenever $u\in D_x$. As a conclusion we get

\begin{theo}\label{theo:diff1}
A $K$-orbit $K.x$ in $M$ is symplectic if and only if the
restriction of the moment map $\mu|_N$ is a diffeomorphism onto a
coadjoint orbit $\mathcal{O}$.
\end{theo}

Suppose now that $N$ defined as above is symplectic. This means that
$K$-action on $N$ is the same as the coadjoint $K$-action on its $\mu$-image
$\mathcal{O}$ (because $\mu$ is a diffeomorphism). Since $K$ is compact there
exists an $\mathrm{Ad}$-invariant scalar product $(\,\cdot\,|\,\cdot\,)$ on
the carrier space $\mathfrak{k}$
\begin{equation}\label{kcprod}
(\mathrm{Ad}(g)X|\mathrm{Ad}(g)Y)=(X|\,Y),\quad \forall
X,Y\in\mathfrak{k},\quad g\in K.
\end{equation}
In particular every operator $\mathrm{Ad}(g)$ is unitary, operators
$\mathrm{ad}_X$ are anitHermitian ($\mathrm{ad}_X^\ast=-\mathrm{ad}_X$), and
\begin{equation}\label{ad}
(\mathrm{ad}_XY|Z)=-(Y|\mathrm{ad}_XZ)
\end{equation}
We may use the invariant scalar product (\ref{kcprod}) to identify
$\mathfrak{k}$ with $\mathfrak{k}^\ast$. More specifically we know
that for any $\alpha\in\mathfrak{k}^\ast$ there exist
$X\in\mathfrak{k}$ such that $\alpha=(X|\,\cdot\,)$. Upon such
identification coadjoint orbits are exactly adjoint ones. To see
this consider $\alpha\in\mathfrak{k}^\ast$. We know that
\begin{equation}\label{k}
\alpha=(X|\,\cdot\,)\equiv\alpha_X,
\end{equation}
for some $X\in \mathfrak{k}$. We need to show that
$\mathrm{Ad}^\ast_g\alpha$ (defined by \ref{Adast}) is equal to
$\alpha_{\mathrm{Ad}(g)X}$. We have
\begin{eqnarray}
\langle\mathrm{Ad}^\ast_g\alpha_X,Y\rangle
=\langle\alpha_X,\mathrm{Ad}(g^{-1})Y\rangle=(X|\mathrm{Ad}(g^{-1})Y)=
\nonumber
\\
=(\mathrm{Ad}(g)X|Y)=\alpha_{\mathrm{Ad}(g)X}, \quad \forall g\in K,
\end{eqnarray}
but this exactly what we wanted. Now we have important fact which
says that adjoint action of $K$ on $\mathfrak{t}$ (the maximal
commutative subalgebra of $\mathfrak{k}$) gives the whole
$\mathfrak{k}$. This observation is true for any compact group but
in the following we will need only its exemplification given by a
familiar example.
\begin{ex}\label{example}
Let $K=SU(n)$ with the Lie algebra $\mathfrak{k}=\mathfrak{su}(n)$ of
traceless antiHermitian matrices. Maximal commutative subalgebra of
$\mathfrak{k}$ consists of traceless diagonal matrices
$\mathfrak{t}=\mathrm{diag}(it_1,\ldots,it_n)$ where $t_k\in R$. It is well
known fact that every antiHermitian matrix has a purely imaginary spectrum
and can be diagonalized by a unitary operator. Therefore, taking any
$X\in\mathfrak{k}$ we can find $U\in U(n)$ such that
\begin{equation}\label{sympproj1}
UXU^{-1}=\mathrm{diag}(it_1,\ldots,it_n)\quad t_k\in\mathbb{R} \quad
\forall k
\end{equation}
Moreover we can choose $SU(n)\ni
U_1=\mathrm{det}(U)^{-\frac{1}{n}}U$ so that $\mathrm{det}(U_1)=1$
and
\begin{equation}\label{sympproj2}
X=U_1^{-1}\mathrm{diag}(it_1,\ldots,it_n)U_1\quad t_k\in\mathbb{R}
\quad \forall k
\end{equation}
Hence indeed, every matrix $X\in\mathfrak{k}$ can be obtained from
the $\mathfrak{t}$ by the adjoint action.
\end{ex}
As a consequence we obtain that every adjoint orbit contains an element of
the maximal commutative subalgebra $\mathfrak{t}$ which is fixed by the
adjoint action of the maximal torus $T\subset G$, where torus $T$ is obtained
by exponentiating $\mathfrak{t}$ ($T=\{e^{t}:t\in\mathfrak{t}\}$). Indeed,
since $T$ is Abelian it fixes its elements by conjugation, $tt^\prime
t^{-1}=t^\prime$, for $t,t^\prime\in T$. By differentiation it translates to
fixing the elements of $\mathfrak{t}$ (and, consequently $\mathfrak{t}^\ast$)
by the adjoint (coadjoint) action of $T$. Combining this observation with
Theorem~\ref{theo:diff1} establishing diffeomorphism of a symplectic orbit
with some coadjoint one, we arrive at the following conclusion

\begin{theo}\label{theo:fixN}
If an orbit $N$ of $K$ through $x\in M$ is symplectic then the set
of points on $N$ fixed by the action of $T$ is nonempty,
$\mathrm{Fix}_N(T)\neq 0$.
\end{theo}

If the point $x\in M$ is fixed by an element $g\in K$ than by the
equivariant property of moment map, its $\mu$-image is also fixed by
adjoint action $\mathrm{Ad}(g)$. The degeneracy subspaces $D_x$
originate from nontrivial action of those symplectomorphisms
$\Phi_g$ for which the corresponding $\mathrm{Ad}(g)$-action on
$\mu(x)$ is trivial. Thus we have following theorem
\cite{kostant82,guillemin84}.

\begin{theo}\label{theo:ks}
The orbit of $K$ through $x\in M$ is symplectic if and only if the
stabilizer subgroup (of the $K$-action) of $x$ is the same as the
stabilizer subgroup (of the $\mathrm{Ad}^\ast$-action) of $\mu(x)$.
\end{theo}

It is always true that
$\mathrm{Stab}(x)\subset\mathrm{Stab}(\mu(x))$ hence,
\begin{Collorary}\label{collo}
The dimension of degeneracy subspace $D_x$ for an orbit $N=K.x$ does
not depend on $x\in N$ and can be computed as
\begin{equation}\label{dimdeg}
D(x)=\mathrm{dim}(D_x)=\mathrm{dim}(\mathrm{Stab}(x))-\mathrm{dim}(\mathrm{Stab}(\mu(x)))=\mathrm{dim}(K.x)-\mathrm{dim}(K.\mu(x)).
\end{equation}
\end{Collorary}
This means we can associate with every orbit of $K$-action an non
negative integer $D(x)$ which measures the degree of its non
symplecticity.

\section{Symplectic orbits in the space of states}
\label{sec:sympstat}

In the case of pure states $M=\mathbb{P}(V)$. The canonical symplectic form
on $\mathbb{P}(V)$, the moment map and symplectic orbits of a unitary $K$
action can be calculated as follows \cite{kostant82,guillemin84}. For
$A\in\mathfrak{u}(V)$ let $A_x\in T_x\mathbb{P}(V)$ be the vector tangent at
$t=0$ to the curve $t\mapsto\pi(\exp(tA)v)$, where $x=\pi(v)$, $v\in V$,
$\|v\|=1$ and $\pi:V\rightarrow \mathbb{P}(V)$ is the canonical projection.
When $A$ runs through the whole Lie algebra $\mathfrak{u}(V)$ the
corresponding $A_x$ span $T_x\mathbb{P}(V)$ and for $A,B\in\mathfrak{u}(V)$
we obtain
\begin{equation}\label{sympproj4}
\omega_x(A_x,B_x)=-\mathrm{Im}\bk{Av}{Bv}=\frac{i}{2}\bk{[A,B]v}{v}.
\end{equation}
The equivariant moment map $\mu:\mathbb{P}(V)\rightarrow
\mathfrak{u}^\ast(V)$ for the action of $U(V)$ on $\mathbb{P}(V)$ is
given by
\begin{equation}\label{momproj}
\mu_A(x)=\frac{1}{2}\bk{v}{Av}.
\end{equation}
The group $K$ acts on $V$ \textit{via} its unitary representation $\varrho
:K\rightarrow U(V)$. The restriction of $\omega$ to $K.x$ can be calculated
as above but now $A$ and $B$ are restricted to elements of $\mathfrak{k}$.
From Section~\ref{sec:symp} we know that the necessary condition for orbit to
be symplectic is possessing a point fixed by the maximal torus $T$ of $K$.
From the definition of weights and weight vectors (\ref{defweight}) it easily
follows that in the case of $K$-action \textit{via} a unitary representation
on a projective space $\mathbb{P}(V)$ fixed points of the $T$ action are
exactly the weight vectors. Hence Theorem~\ref{theo:fixN} can be reformulated
as,

\begin{fact}
Let $K$ act on $\mathbb{P}(V)$ by unitary representation on a Hilbert space
$V$. If $N=K.x$ is a symplectic $K$-orbit then $N$ contains a point
$x=\pi(v)$ where $v$ is a $T$-weight vector, i.e. $v\in V_\lambda$ fore some
weight $\lambda$.
\end{fact}

Our goal is to find a sufficient condition for $N=K.x$ to be
symplectic. This condition is of course given in
Theorem~\ref{theo:ks} but we want to have it in more useful form. It
is enough to restrict our attention to orbits passing through the
weight vectors. For $v\in V_\lambda$ we consider the tangent space
$T_x(N)$ equipped with the 2-form $\omega_x$. Let $\alpha$ be a
positive root and define $\mathcal{O}_\alpha$ to be the orbit of
$SU_2(\alpha)$ of the associated $SU_2$-triple. Let $P_\alpha$
denote the tangent space to $\mathcal{O}_\alpha$ at the point $x$.
Tangent space $T_xN$ can be of course considered as the collection
of $P_\alpha$ where $\alpha$ range over all positive roots. Let
$\mathfrak{k}^{\mathbb{C}}$ be the complexification of
$\mathfrak{k}$ - the Lie algebra of $K$. It has the root-space
decomposition
\begin{equation}\label{rootcompl}
\mathfrak{k}^{\mathbb{C}}=\mathfrak{t}^{\mathbb{C}} \mathop  \oplus
\limits_\alpha \mathbb{C}E_\alpha,
\end{equation}
where $E_\alpha$ is a root vector corresponding to the root
$\alpha$, hence
$[E_\alpha,E_{-\alpha}]=H_\alpha\in\mathfrak{t}^{\mathbb{C}}$. The
corresponding decomposition of $\mathfrak{k}$ reads
\begin{equation}\label{rootreal}
 \mathfrak{k}=\mathfrak{t}\mathop  \oplus \limits_\alpha
\mathbb{R}\left(E_\alpha-E_{-\alpha}\right)\mathop  \oplus
\limits_\alpha \mathbb{R}i\left(E_\alpha+E_{-\alpha}\right),
\end{equation}
where $\alpha$ ranges over all positive roots.
\begin{fact} If $\alpha$ and $\beta$ are different positive roots,
then the tangent planes $P_\alpha$ and $P_\beta$ are
$\omega$-orthogonal.
\end{fact}
$Proof.$ The symplectic form $\omega_x$ is given as:
\begin{equation}\label{omegax}
\omega_x(A_x,B_x)=\frac{i}{2}\bk{[A,B]v}{v}.
\end{equation}
We have assumed that $v\in V_\lambda$. If $[A,B]v$ is in some other
weight space, the right hand side of (\ref{omegax}) vanishes since
two different weight spaces are orthogonal. We know that $P_\alpha =
\mathrm{Span}\{(E_\alpha-E_{-\alpha}).v, i(E_\alpha+ E_{-\alpha}).v
\}$ and $P_\beta=\mathrm{Span}\{(E_\beta-E_{-\beta}).v,
i(E_\beta+E_{-\beta}).v\}$. We also  know that
$[E_\alpha,E_\beta].v\in V_{\lambda+\alpha+\beta}$ or is equal zero.
But $V_{\lambda+\alpha+\beta}$ is orthogonal to $V_{\lambda}$.
Consequently, if $A_x\in P_\alpha$ and $B_x\in P_\beta$ then
\begin{equation}\label{omegax0}
\omega_x(A_x,B_x)=\frac{i}{2}\bk{[A,B]v}{v}=0,
\end{equation}
which is what we wanted to prove.
\bigskip

\indent Summing up we know that $T_x N=\bigcup_\alpha P_\alpha$ and
that spaces $P_\alpha$ are $\omega$-orthogonal. So $T_x N$ is
symplectic vector space if and only if all $P_\alpha$ are
symplectic.
\begin{fact}
The space $P_\alpha$ is symplectic if and only if
$\bk{[E_\alpha,E_{-\alpha}]v}{v}\neq 0$
\end{fact}
$Proof.$ Let $A_x\in P_\alpha$ and $B_x\in P_\alpha$. Computing
\begin{equation}\label{omegax1}
\omega_x(A_x,B_x)=\frac{i}{2}\bk{[A,B]v}{v}
\end{equation}
we see that only the term $\bk{[E_{\alpha},E_{-\alpha}]v}{v}$ can
give a nonzero result and when it indeed does not vanish then
$P_\alpha$ is symplectic which is what we wanted to prove.
\bigskip

So $T_x N$ is symplectic when the following implication is true
\begin{equation}\label{implication}
\bk{[E_\alpha,E_{-\alpha}]v}{v}= 0 \Rightarrow P_\alpha =0.
\end{equation}
The left hand side of (\ref{implication}) can be rewritten as
\begin{eqnarray}
[E_{\alpha},E_{-\alpha}]v=H_\alpha .v=\lambda(H_\alpha)v,
\end{eqnarray}
where $\lambda$ is the weight of $v$. For the right hand side of
(\ref{implication}) recall that $P_\alpha =
\mathrm{Span}\{(E_\alpha-E_{-\alpha}).v, i(E_\alpha+ E_{-\alpha}).v \}$. So
$P_\alpha=0$ means $E_\alpha v=0=E_{-\alpha}v$. Hence \cite{kostant82},
\begin{theo}[Kostant-Sternberg]\label{theo:KoSte}
The orbit $N=K.x$, $x=\pi(v)$, $v\in V_\lambda$ for some weight
$\lambda$, is symplectic if and only if for every positive root
$\alpha$ with $\lambda(H_\alpha)=0$ it follows that $E_\alpha
v=0=E_{-\alpha}v$.
\end{theo}

To demonstrate how this theorem works we will prove that the orbit
through the highest weight vector is always symplectic. As it was
mentioned in Subsection \ref{generalcase} every unitary irreducible
representation of compact semisimple group $K$ is highest weight
representation. The highest weight vector is defined as follows
\begin{defi}\label{highestweight}
Let $K$ be compact semisimple Lie group and denote by $\mathfrak{k}$
its Lie algebra and by $\mathfrak{k}^\mathbb{C}$ its
complexification. Then $\mathfrak{k}^\mathbb{C}$ admits
decomposition (\ref{rootcompl}). The weight vector $v\in V_\lambda$
of irreducible representation of $K$ (respectively $\mathfrak{k}$ or
$\mathfrak{k}^\mathbb{C}$) is highest weight if and only if
\begin{equation}\label{highest}
    E_\alpha.v=0,
\end{equation}
where $\alpha$ range through all positive roots.
\end{defi}
Let us take $v\in V_\lambda$ - the highest weight vector of irreducible
representation of $\mathfrak{k}^\mathbb{C}$ and consider corresponding orbit
$N=K.x$, where $x=\pi(v)$. It is easy to see that according to Definition
\ref{highestweight}, $v$ is also the highest weight vector for all
$\mathfrak{sl}_2(\alpha)$-triples. From the representation theory we know
that weights of irreducible representation of $\mathfrak{sl}_2(\alpha)$ are
$W=\{-n,-n+2,\ldots,n-2,n\}$, where $n\geq 0$. This means
\begin{equation}
    H_\alpha.v=\lambda(H_\alpha)v=nv
\end{equation}
So $\lambda(H_\alpha)=n$. The only interesting $\alpha$ is the one for which
$n=0$. But then we have one dimensional, hence trivial, representation of
$sl_2(\alpha)$. This, of course, means $E_{-\alpha}.v=0=E_{\alpha}.v$. Making
use of Theorem \ref{theo:KoSte} we see that $N$ is symplectic. In fact this
orbit is not only symplectic but also K\"{a}hler (see Appendix for definition
of a K\"{a}hler manifold). Indeed, from Theorem \ref{apptheo} we know that to
prove this, it is enough to check that this orbit is complex manifold since
$\mathbb{P}(V)$ is positive K\"{a}hler manifold (see Appendix). But if $v$ is
the highest weight vector then the tangent space $T_xN=\bigcup_\alpha
P_\alpha$, where $\alpha$ range over positive roots and
$P_\alpha=\mathrm{Span}\{E_{-\alpha}.v,iE_{-\alpha}.v\}$. So $T_xN$ is stable
under multiplication by $i$ hence $N$ is complex.

\section{Distinguishable particles}
\subsection{Two qubits case}
In the simplest case of two qubits we may use directly the
Kostant-Sternberg theorem from the last section. The Hilbert space
is then $\mathcal{H} =\mathbb{C}^2\otimes\mathbb{C}^2$  and the
direct product $K=SU(2)\times SU(2)$ acts on $\mathcal{H}$ in a
natural way,
\begin{equation}\label{actionprod}
(g_1,g_2)v_1\otimes v_2=g_1v_1\otimes g_2 v_2,
\end{equation}
where $g_1,g_2\in SU(2)$ and $v_1,v_2\in \mathbb{C}^2$. Our first
goal is to identify symplectic orbits of $K$. To apply theorems and
facts established in the previous sections we start with the
root-space decomposition of the Lie algebra
$\mathfrak{k}^{\mathbb{C}}=\mathfrak{sl}(2,\mathbb{C})\oplus
\mathfrak{sl}(2,\mathbb{C})$, i.e., the complexification of
$\mathfrak{k}=\mathfrak{su}(2)\oplus\mathfrak{su}(2)$ - the Lie
algebra of $G$. The algebra $\mathfrak{k}^{\mathbb{C}}$ is
semisimple as a direct sum of simple algebras
$\mathfrak{sl}(2,\mathbb{C})$. Let us remind that,
\begin{eqnarray}
\mathfrak{sl}(2,\mathbb{C})=\mathrm{Span}\{X,H,Y\},
\\ \nonumber
[H,X]=2X,\quad [H,Y]=-2Y,\quad [X,Y]=H.
\end{eqnarray}
The Cartan subalgebra of $\mathfrak{sl}(2,\mathbb{C})$ is spanned by
$H$, whereas $\mathrm{Span}\{X\}$, $\mathrm{Span}\{Y\}$ are the
positive and negative root spaces. An element of
$\mathfrak{k}^{\mathbb{C}}$ can be written as $(Z_1,Z_2)$, where
$Z_1,Z_2 \in \mathfrak{sl}(2,\mathbb{C})$. We also have:
\begin{eqnarray}
[(Z_1,Z_2),(W_1,W_2)]=([Z_1,W_1],[Z_2,W_2]).
\end{eqnarray}
Knowing this we find that the Cartan subalgebra of
$\mathfrak{k}^{\mathbb{C}}$ is
$\mathfrak{t}=\mathrm{Span}\{(H,0),(0,H)\}$. The commutation
relations read as,
\begin{alignat}{3}
[(H,0),(X,0)]&=2(X,0),\quad [(H,0),(Y,0)]&=-2(Y,0),\quad
[(X,0),(Y,0)]&=(H,0),
\nonumber \\
[(0,H),(0,X)]&=2(0,X),\quad [(0,H),(0,Y)]&=-2(0,Y),\quad
[(0,X),(0,Y)]&=(0,H),
\nonumber \\
[(0,W),(Z,0)]&=0.
\end{alignat}
Since $\mathfrak{k}^{\mathbb{C}}$ is semisimple, its root spaces are
one dimensional.  We have the following roots (computed in the basis
$\{(H,0),(0,H)\}$ of the Cartan subalgebra $\mathfrak{t}$), and the
corresponding root spaces,
\begin{alignat}{2}
 \alpha & =\phantom{-}(2,0),   &  \qquad  V_\alpha\phantom{_-} &
=\mathrm{Span}\{(X,0)\}, \\
-\alpha & =(-2,0),  &  \qquad  V_{-\alpha} &
=\mathrm{Span}\{(Y,0)\}, \\
 \beta  & =\phantom{-}(0,2),   &  \qquad  V_\beta\phantom{_-} &
=\mathrm{Span}\{(0,X)\}, \\
-\beta  & =(0,-2),  &  \qquad  V_{-\beta}  &
=\mathrm{Span}\{(0,Y)\}.
\end{alignat}
Thus we have the following decomposition of
$\mathfrak{k}^{\mathbb{C}}$,
\begin{eqnarray}
\mathfrak{k}^{\mathbb{C}}&=&\mathfrak{n}_-\oplus\mathfrak{t}
\oplus\mathfrak{n}_+\, ,\\
\mathfrak{n}_-&=&\mathrm{Span}\{(Y,0),(0,Y)\},\\
\mathfrak{n}_+&=&\mathrm{Span}\{(X,0),(0,X)\},\\
\mathfrak{t}\phantom{_+}&=&\mathrm{Span}\{(H,0),(0,H)\},
\end{eqnarray}
where $\mathfrak{n}_-$ and $\mathfrak{n}_+$ are negative and
positive root spaces, respectively. Let
\begin{equation}
e_1=\left(
  \begin{array}{c}
    1 \\
    0 \\
  \end{array}
\right),\quad e_2=\left(
             \begin{array}{c}
               0 \\
               1 \\
             \end{array}
           \right),
\end{equation}
be the standard basis of $\mathbb{C}^2$. The Lie algebra
$\mathfrak{sl}(2,\mathbb{C})$ acts then {\it via} the defining
representation,
\begin{equation}
H=\left(
    \begin{array}{cc}
      1 & 0 \\
      0 & -1 \\
    \end{array}
  \right),\quad X=\left(
                   \begin{array}{cc}
                     0 & 1 \\
                     0 & 0 \\
                   \end{array}
                 \right),
  \quad Y=\left(
            \begin{array}{cc}
              0 & 0 \\
              1 & 0 \\
            \end{array}
          \right).
\end{equation}
The highest weight vector equals $e_1$ and there are just two weight
spaces, one spanned by $e_1$ and the other by $e_2$. The
corresponding weights are $1$ and $-1$.

The action of $(Z_1,Z_2)\in \mathfrak{k}^{\mathbb{C}}$ on
$\mathcal{H}$ is given by
\begin{equation}
(Z_1,Z_2)v_1\otimes v_2=Z_1v_1\otimes v_2 + v_1\otimes Z_2 v_2.
\end{equation}
It is easy to guess that the highest weight vector for the above
representation equals $e_1\otimes e_1$.  Indeed, it is an
eigenvector of the Cartan subalgebra and is annihilated by all
elements of $\mathfrak{n}_+$. The weight spaces are obtained by
successive action of $\mathfrak{n}_-$ on $e_1\otimes e_1$. In the
basis $\{(H,0),(0,H)\}$ the weights and weight vectors read as,
\begin{alignat}{3}
\lambda_1&=(1,1), &\quad & v_1=e_1\otimes e_1, \\
\lambda_2&=(1,-1), &\quad & v_2=e_1\otimes e_2, \\
\lambda_3&=(-1,1), &\quad & v_3=e_2\otimes e_1, \\
\lambda_4&=(-1,-1),&\quad & v_4=e_2\otimes e_2.
\end{alignat}
The $\mathfrak{sl}(2,\mathbb{C})$ triples corresponding to the
positive roots of $\mathfrak{k}^{\mathbb{C}}$ are
$\{(X,0),(H,0),(Y,0)\}$ and $\{(0,X),(0,H),(0,Y)\}$. To decide if an
orbit through a weight vector is symplectic it is enough to check if
$\lambda((0,H))\neq 0$ and $\lambda((H,0)))\neq 0$ where $\lambda$
is one of the weights from the list above. Since weights are given
by two non-zero numbers $(n_1,n_2)$, we find
\begin{fact}
In the case of two qubits, only the orbits through weight vectors are
symplectic in the projective space
$\mathbb{P}(\mathbb{C}^2\otimes\mathbb{C}^2)$. In fact all weight vectors lie
on the same orbit which is K\"ahler and contains all separable states. Orbits
through entangled states are not symplectic
\end{fact}
Let us now consider the states $e_1\otimes e_2\pm e_2\otimes e_1$
which are not weight vectors and are not separable. The orbits
through them are not symplectic and we can ask what is the dimension
of the degeneracy subspace for them. We need to examine which
vectors from the tangent space to the orbit of $SU(2)\times SU(2)$
are tangent to the fibers of the corresponding moment map $\mu$. We
already know that,
\begin{eqnarray}
\mu_A(x)=\frac{1}{2}\bk{v}{Av},\quad x=\pi (v).
\end{eqnarray}
In our case we have,
\begin{eqnarray}
\mu_{(Z_1,Z_2)}(x)=\frac{1}{2}\bk{e_1\otimes e_2\pm e_2\otimes
e_1}{(Z_1,Z_2)(e_1\otimes e_2\pm e_2\otimes
e_1)}= \nonumber\\
=\mathrm{tr}(Z_1)+\mathrm{tr}(Z_2)=0, \quad (Z_1,Z_2)\in
\mathfrak{k},
\end{eqnarray}
where the first equality was obtained by a direct computation and
the second one is a consequence of the zero-trace property of the
matrices from $\mathfrak{su}(2)$. Thus the degeneracy space is the
whole tangent space to the orbit through state $e_1\otimes e_2\pm
e_2\otimes e_1$. This space can be directly computed, as it is
spanned by the projection of vectors given by
\begin{eqnarray}
(Z_1,Z_2)(e_1\otimes e_2\pm e_2\otimes e_1)=Z_1e_1\otimes
e_2+e_1\otimes Z_2e_2\pm Z_1e_2\otimes e_1\pm e_2\otimes Z_2e_1.
\end{eqnarray}
Using the Pauli matrices multiplied by the imaginary unit $i$ as a basis for
$\mathfrak{su}(2)$ and the formula (57) we obtain that in both cases the
tangent space is three dimensional. In the case of the state $e_1\otimes e_2+
e_2\otimes e_1$ it is spanned by $\{i(e_1\otimes e_1 +e_2\otimes
e_2),(e_2\otimes e_2 - e_1\otimes e_1), i(e_1\otimes e_2 - e_2\otimes
e_1)\}$, whereas for the state $e_1\otimes e_2-e_2\otimes e_1$ it is spanned
by $\{i(e_2\otimes e_2 -e_1\otimes e_1),(e_1\otimes e_1+e_2\otimes
e_2),i(e_1\otimes e_2 +e_2\otimes e_1)\}$. The conclusion is that we can use
the dimension of the degeneracy space as a measure of entanglement.

In principle a similar reasoning directly using the
Kostant-Sternberg theorem can be applied in cases of larger
dimensions of subsystems an/or for many-partite systems involving
multiple tensor products of spaces with arbitrary dimensions, but
explicit calculations become prohibitively complicated. In the next
section we present a method allowing for finding the degeneracy
spaces for bipartite systems of arbitrary dimensions based on the
Singular Value Decomposition (SVD) of a matrix, and in the following
one we show how to extend the reasoning to a multipartite case where
the direct application of SVD is not possible.

\section{Degeneracy subspaces and SVD}
The method of determining the dimension of the degeneracy space
presented in the previous section can be extended to a more general
case of two distinguishable particles, but in this case one can
achieve the goal in a less cumbersome manner by invoking the
Singular Value Decomposition of an arbitrary complex matrix
\cite{horn85}. We will present the solution for two distinguishable
but otherwise identical particles (i.e.\ living in spaces of the
same dimension $N$). A generalization to unequal dimensions of the
spaces needs only a little bit more effort.

The Hilbert space is thus now $\mathcal{H}=\mathbb{C}^N\otimes
\mathbb{C}^N$. Let us fix an orthonormal basis $\{e_i:\, i=1,\ldots
,N\}$ of $\mathbb{C}^N$. (e.g., the standard one where $e_i$ is a
column vector with one on the $i$-th position and zero on others).
Any state $\ket{\Psi}\in \mathcal {H}$ can be decomposed as:
\begin{equation}\label{bistate}
\ket{\Psi}=\sum_{i,j=1}^{N}C_{ij}e_i\otimes e_j
\end{equation}
The action of $U\otimes V\in SU(N)\times SU(N)$ gives:
\begin{eqnarray}\label{svd1}
U\otimes V\ket{\Psi}=\sum_{i,j=1}^{N}C_{ij}Ue_i\otimes
Ve_j=\sum_{i,j=1}^{N}C_{ij}U_{ki}e_k\otimes V_{lj}e_l
=\sum_{k,l=1}^{N}(UCV^T)_{kl}e_k\otimes e_l
\end{eqnarray}
It is well known fact that any complex matrix can be put to a
diagonal form by the simultaneous left and right action of the
unitary group achieving the SVD, i.e.\ there exist unitary
$\tilde{U},\tilde{V}$ such that
\begin{equation}\label{svd2}
\tilde{U}C\tilde{V}=\mathrm{diag}(0,\ldots,0,\nu_1,\ldots,\nu_2,\ldots,\nu_K),
\end{equation}
where $\nu_i>0$ and
$\{0,\ldots,0,\nu_1^2,\ldots,\nu_2^2,\ldots,\nu_K^2\}$ constitute
the spectrum of $C^\dagger C$ (and, equivalently, the spectrum of
$CC^\dagger$). Taking $U=\tilde{U}$ and $V=\tilde{V}^T$ in
(\ref{svd1}) we conclude that the orbit of $SU(N)\times SU(N)$
through any state $\ket{\Psi}$ contains a point which can be written
as:
\begin{equation}\label{svd3}
\ket{\Psi^\prime}=\sum_{i=1}^N  p_i e_i\otimes e_i,
\end{equation}
with $ p_i\ge 0$ and $\sum_{i=1}^N  p_i^2=1$.

We denote by $m_i$ the multiplicity of $\nu_i$ and by $m_0$ the
dimension of the kernel of $C$, hence $m_0+\sum_{n=1}^K m_i=N$. We
can use state $\ket{\Psi^\prime}$ to compute the dimension of orbit
through $\ket{\Psi}$. The crucial for this is the observation
\cite{szk02} that $\ket{\Psi^\prime}$ is stabilized by the action of
$U\otimes V$ where:
\begin{equation}\label{stabUV}
U=\left(
  \begin{array}{cccc}
    u_0 & &  &  \\
     & u_1 &  &  \\
     &  & \ddots &  \\
     &  &  & u_K \\
  \end{array}
\right),V=e^{i\phi}\left(
          \begin{array}{cccc}
            v_0 &  &  &  \\
             & \overline{u}_1 &  &  \\
             &  & \ddots & \\
             &  &  & \overline{u}_K \\
          \end{array}
        \right),
\end{equation}
$u_1,\ldots,u_K$ are arbitrary unitary operators from, respectively,
$U(m_1),\ldots,U(m_K)$. Both $u_0$  and $v_0$ belong to $U(m_0)$ and
$\mathrm{det}(u_0)$ and $\mathrm{det}(v_0)$ are fixed by the determinants
$u_1,\ldots,u_K$ in a way ensuring that matrices $U,V$ are special unitary.
Knowing this we can compute the dimension of the orbit $\mathcal{O}$ of
$G=SU(N)\times SU(N)$ through $\ket{\Psi}$ in the projective space
$\mathbb{P}(\mathcal{H})$ as:
\begin{equation}\label{orbit:proj}
\mathrm{dim}(\mathcal{O})=(2N^2-2)-\big((2m_0^2-2)+1+\sum_{n=1}^K
m_n^2 \big)=2N^2-2m_0^2-\sum_{n=1}^K m_n^2-1,
\end{equation}
where we used $\mathrm{dim}(U(n))=n^2=\mathrm{dim}(SU(n))+1$. The
dimensions of the two $U(m_0)$ blocks are diminished by one due to
the determinant fixing condition stated above, and an additional one
is subtracted due to the projection on $\mathbb{P}(\mathcal{H})$.

To compute the dimension of the coadjoint orbit in the dual space to
$\mathfrak{k}=\mathfrak{su}(N)\oplus \mathfrak{su}(N)$ associated
with $\ket{\Psi^\prime}$ via the moment map $\mu$ let us calculate
\begin{eqnarray}\label{svd4}
\mu_{(A,B)}(\sum_{i=1}^N  p_i e_i\otimes e_i)=\bk{\sum_{i=1}^N
 p_i e_i\otimes e_i}{(A\otimes I+I\otimes B)(\sum_{j=1}^N
 p_j e_j\otimes e_j)}= \nonumber \\
 =\sum_{i,j=1}^N\bk{p_ie_i\otimes
e_i}{p_iAe_j\otimes e_j+ p_je_j\otimes Be_j}=\sum_{i=1}^N
 p_i^2\big(\bk{e_i}{Ae_i}+\bk{e_i}{Be_i}\big),
\end{eqnarray}
with $(A,B)\in \mathfrak{su}(N)\oplus\mathfrak{su}(N)$. It is easy
to see (using the standard basis of $\mathfrak{su}(N)$) that in fact
$SVD$ transfers our state $\ket{\Psi}$ into a state
$\ket{\Psi^\prime}$ such that $\mu(\ket{\Psi^\prime})\in
\mathfrak{t}^\ast$, where $\mathfrak{t}^\ast$ is the dual space to
the Cartan subalgebra $\mathfrak{t}$ of $\mathfrak{k}$. Of course
every coadjoint orbit is passing through at least one point from
$\mathfrak{t}^\ast$  (usually, a coadjoint orbit contains more than
one point from $\mathfrak{t}^\ast$, all this points lie on the orbit
of the Weyl group). This fact could be seen as a geometrical
interpretation of the SVD, but in contrast to the SDV itself, it
remains true in multipartite cases. We will use it in the following
sections. Going back to our considerations we know that
$\mu(\ket{\Psi^\prime})$ is determined by the action on
$\mathfrak{t}$ which is generated by $I\otimes H,H\otimes I$ where
$H$ belongs to the Cartan subalgebra of $\mathfrak{su}(N)$. Using
invariant scalar product on $\mathfrak{k}$ given by $\tr{AB}$, we
can find an element $(X,Y)\in\mathfrak{t}$ such that
$\mu(\ket{\Psi^\prime})=\alpha_{(X,Y)}$. To compute this we use
standard basis for $\mathfrak{t}$ given by
$H_1=\mathrm{diag}(1,-1,0,\ldots,0),\ldots,H_{N-1}=\mathrm{diag}(0,0,\ldots,1,-1)$.
We have
\begin{eqnarray}
\mu_{(H_k,0)}(\ket{\Psi^\prime})=\mathrm{tr}((X\otimes I+I\otimes
Y)H_k\otimes I)=\mathrm{tr}(XH_k), \nonumber \\
\mu_{(0,H_k)}(\ket{\Psi^\prime})=\mathrm{tr}((X\otimes I+I\otimes
Y)I\otimes H_k)=\mathrm{tr}(YH_k),
\end{eqnarray}
but we also know that
\begin{equation}
\mu_{(H_k,0)}(\ket{\Psi^\prime})=\sum_{i=1}^N
p_i^2(e_i|H_ke_i)=p_k^2-p_{k+1}^2=\mu_{(0,H_k)}(\ket{\Psi^\prime}).
\end{equation}
It is easy to see now that $X=Y$ and
\begin{equation}\label{XY}
X=Y=\mathrm{diag}(-\frac{1}{N}+p_1^2,-\frac{1}{N}+p_2^2,\ldots,
-\frac{1}{N}+p_N^2).
\end{equation}
To compute the dimension of the coadjoint orbit through
$\mu(\ket{\Psi^\prime})$ notice that if $U_1\otimes U_2\in G$ then
\begin{equation}
U_1\otimes U_2(X\otimes I+I\otimes X)U_1^\dagger\otimes
U_2^\dagger=U_1XU_1^\dagger\otimes I+I\otimes U_2XU_2^\dagger.
\end{equation}
Hence, to obtain the dimension of the coadjoint orbit we need to
compute the dimension of the stabilizer subgroup of $X$ by the
adjoint action. It is easy to see that $X$ is stabilized by any
matrix of the form:
\begin{equation}
U=\left(
  \begin{array}{cccc}
    u_0 &  &  &  \\
     & u_1 &  &  \\
     &  & \ddots &  \\
     &  &  & u_K \\
  \end{array}
\right)
\end{equation}
where $u_0,u_1,\ldots,u_K$ are arbitrary unitary operators from
$U(m_0),U(m_1),\ldots,U(m_K)$ and the value of $\mathrm{det}(u_0)$
is fixed by demanding that $U$ is special unitary. The dimension of
the coadjoint orbit through $\mu(\ket{\Psi^\prime})$ is thus
\begin{equation}
\mathrm{dim}(\mu(\mathcal{O}))=(2N^2-2)-(2\sum_{n=0}^K
m_n^2-2)=2N^2-2\sum_{n=0}^Km_n^2
\end{equation}
Now we are able to compute the dimension $D(\ket{\Psi})$ of the
degeneracy subspaces (fibers of the moment map),
\begin{equation}\label{degqubit}
D(\ket{\Psi})=\mathrm{dim}(\mathcal{O})-\mathrm{dim}(\mu(\mathcal{O}))
=\sum_{n=1}^Km_n^2-1,
\end{equation}
and we see that the orbit through $\ket{\Psi}$ is symplectic if and
only if in the $SVD$ decomposition we get diagonal matrix with only
one non zero entry.
\begin{fact}
In the case of two identical but distinguishable particles there is only one
symplectic orbit in the projective space
$\mathbb{P}(\mathbb{C}^N\otimes\mathbb{C}^N)$. This orbit contains all
separable states and is K\"ahler. Orbits through entangled states are not
symplectic.
\end{fact}
Knowing this and making use of Corollary \ref{collo} we arrive with
\begin{fact}
In the case of two identical but distinguishable particles the dimension of
the degeneracy space $D(\ket{\Psi})=\sum_{n=1}^Km_n^2-1$ gives a well defined
entanglement measure.
\end{fact}

\section{Three particle case} \label{sec:3part}

As already mentioned the SVD has no generalization to multiple
tensor products corresponding to multiparticle cases. Nevertheless
we may apply some methods from the previous section if we look at
the SVD from a slightly different point of view. Let us namely ask
the question about necessary conditions for a state $\ket{\Psi}$
(\ref{bistate}) to be sent by the moment map $\mu$ to an element of
$\mathfrak{t}^{\ast}$ represented by $X\otimes I+I\otimes
Y\in\mathfrak{t}$ upon the identification of $\mathfrak{t}^\ast$ and
$\mathfrak{t}$ through the invariant scalar product on
$\mathfrak{k}$. We have,
\begin{gather*}
\mu_{(A,B)}(\ket\Psi)=\bk{\sum_{i,j=1}^{N}C_{ij}e_{i}\otimes e_{j}}
{(A\otimes I+I\otimes B)\sum_{m,n=1}^{N}C_{mn}e_{m}\otimes e_{n}}= \\
\sum_{m,n=1}^{N}\sum_{i,j=1}^{N}\bar{C}_{ij}C_{mn}\big(\delta_{nj}
\bk{e_{i}}{Ae_{m}}+\delta_{im}\bk{e_{j}}{Be_{n}}\big)=
\sum_{i,j,m=1}^{N}\bar{C}_{ij}C_{mj}\bk{e_{i}}{Ae_{m}}+ \\
+\sum_{i,j,n=1}^{N}\bar{C}_{ij}C_{in}\bk{e_{j}}{Be_{n}}=
\sum_{i,m=1}^{N}(CC^{\dagger})_{mi}\bk{e_{i}}{Ae_{m}}
+\sum_{j,n=1}^{N}(C^{\dagger}C)_{jn}\bk{e_{j}}{Be_{n}}=\\
=\sum_{i,j=1}^{N}(CC^{\dagger})_{ji}\bk{e_{i}}{Ae_{j}}+
(C^{\dagger}C)_{ji}\bk{e_{j}}{Be_{i}}.
\end{gather*}
In the following we will denote by $E_{ij}$ the matrix with zero
entries everywhere except $1$ on the $(i,j)$ position. Matrices
$i(E_{ij}+E_{ji})$ and $E_{ij}-E_{ji}$ supplemented by the
previously defined standard basis elements of $\mathfrak{t}$
constitute a standard basis of $\mathfrak{k}$. Taking now $A$ and
$B$ of the form $i(E_{ij}+E_{ji})$ and $E_{ij}-E_{ji}$ which do not
belong to $\mathfrak{t}$ but are from $\mathfrak{k}$, we must have
\begin{gather*}
\sum_{i,j=1}^{N}(CC^{\dagger})_{ji}(e_{i}|(E_{kl}+E_{lk})e_{j})=0,\\
\sum_{i,j=1}^{N}(CC^{\dagger})_{ji}(e_{i}|(E_{kl}-E_{lk})e_{j})=0,
\end{gather*}
and the same for $C^{\dagger}C$. Notice that,
\begin{gather}\label{oofd}
\bk{e_{i}}{(E_{kl}\pm E_{lk})e_{j}}
=\delta_{lj}(e_{i}|e_{k})\pm\delta_{kj}(e_{i}|e_{l})
=\delta_{lj}\delta_{ik}\pm\delta_{kj}\delta_{il},
\end{gather}
hence,
\begin{gather*}
(CC^{\dagger})_{lk}+(CC^{\dagger})_{kl}=0,\\
(CC^{\dagger})_{lk}-(CC^{\dagger})_{kl}=0,
\end{gather*}
and the same equations are fulfilled by $C^{\dagger}C$. It means
that both $CC^{\dagger}$ and $C^{\dagger}C$ are diagonal. From
linear algebra we know that the spectra of $(CC^\dagger)$ and
$(C^\dagger C)$ are the same. Using this property and an additional
freedom of a unitary action which permutes elements on the diagonal
we notice that it is always possible to have
$CC^{\dagger}=C^{\dagger}C$ (i.e. $C$ is a normal operator) and thus
$X=Y$ in $X\otimes I+I\otimes Y$ for corresponding image of moment
map. Let thus
\begin{gather*}
C^{\dagger}C=\left(\begin{array}{cccc}
0 I_{m_0}\\
 & v_{1}I_{m_1}\\
 &  & \ddots\\
 &  &  & v_{K}I_{m_k}\end{array}\right),
\end{gather*}
where $I_n$ is the unit $n\times n$ matrix. Then,
\begin{gather}\label{C}
C=\left(\begin{array}{cccc}
0\\
 & \sqrt{v_{1}}u_1\\
 &  & \ddots\\
 &  &  & \sqrt{v_{K}}u_{K}\end{array}\right),
\end{gather}
where $m_{1},\ldots,m_{k}$ are dimensions of degeneracy of
$v_{1},\ldots,v_{K}$, respectively and $u_n$ are $m_{n}\times m_{n}$ unitary
matrices. Among all matrices (\ref{C}) there is one which is diagonal and
corresponds to the SVD. In this way we proved the existence of the SVD for
any state using the fact that each adjoint orbit intersects the Cartan
subalgebra $\mathfrak{t}$. The second important observation is that all
states (\ref{C}) are sent by the moment map into the same point $X\otimes
I+I\otimes X$ and therefore constitute the fiber of the moment map. The
dimension is of this fiber is $\sum_{n=1}^Km_n^2-1$ which is exactly
$D(\ket{\Psi})$ from the previous section.

To a certain point we may repeat the reasoning in a multipartite case. Thus,
e.g. for a general three-particle state
\begin{gather}
\ket\Psi=\sum_{i,j,k=1}^{N}C_{ijk}e_{i}\otimes e_{j}\otimes e_{k},
\end{gather}
the action of the moment map on $\ket\Psi$ gives:
\begin{gather*}
\mu_{(A,B,D)}(\ket\Psi)= \\
\bk{\sum_{i,j,k=1}^{N}C_{ijk}e_{i}\otimes e_{j} \otimes
e_{k}}{(A\otimes I\otimes I+I\otimes B\otimes I+I\otimes I\otimes
D)\sum_{m,n,l=1}^{N}C_{mnl}e_{m}\otimes
e_{n}\otimes e_{l}}=\\
\sum_{m,n,l=1}^{N}\sum_{i,j,k=1}^{N}\bar{C}_{ijk}C_{mnl}\bk{e_{i}\otimes
e_{j}\otimes e_{k}}{Ae_{m}\otimes e_{n}\otimes e_{l}+e_{m}\otimes
Be_{n}\otimes e_{l}+e_{m}\otimes
e_{n}\otimes De_{l}}=\\
\sum_{m,n,l=1}^{N}\sum_{i,j,k=1}^{N}\bar{C}_{ijk}C_{mnl}
\big(\delta_{jn}\delta_{kl}\bk{e_{i}}{Ae_{m}}+\delta_{im}\delta_{kl}
\bk{e_{j}}{Be_{n}}+\delta_{im}\delta_{jn}\bk{e_{k}}{De_{l}}\big)=\\
\sum_{i,j,k,m=1}^{N}\bar{C}_{ijk}C_{mjk}\bk{e_{i}}{Ae_{m}}
+\bar{C}_{ijk}C_{imk}\bk{e_{j}}{Be_{m}}
+\bar{C}_{ijk}C_{ijm}\bk{e_{k}}{De_{m}}).
\end{gather*}
Again, we want to find conditions for $C_{ijk}$ under which
$\mu_{(A,B,D)}(\ket\Psi)$ belongs to $\mathfrak{t}^{\ast}$.
Substituting for $A$, $B$, and $D$ basis elements from
$\mathfrak{k}-\mathfrak{t}$ and again using (\ref{oofd}) we get:
\begin{gather*}
\sum_{j,k=1}^{N}\bar{C}_{njk}C_{ljk}+
\sum_{j,k=1}^{N}\bar{C}_{ljk}C_{njk}=0, \\
\sum_{j,k=1}^{N}\bar{C}_{njk}C_{ljk}-
\sum_{j,k=1}^{N}\bar{C}_{ljk}C_{njk}=0,
\end{gather*}
and similar two pairs of equations for other combination of indices.
If we now define
\begin{gather}
(C^{1})_{nl}=\sum_{j,k=1}^{N}\bar{C}_{njk}C_{ljk}, \nonumber \\
(C^{2})_{nl}=\sum_{j,k=1}^{N}\bar{C}_{jnk}C_{jlk}, \label{Cs} \\
(C^{3})_{nl}=\sum_{j,k=1}^{N}\bar{C}_{jkn}C_{jkl},\nonumber
\end{gather}
the obtained conditions mean that the matrices $C^{1},C^{2},C^{3}$
are diagonal. In this case it is not generally true that
$C^{1}=C^{2}=C^{3}$ so the corresponding state in $\mathfrak{t}$ is
$X\otimes I\otimes I+I\otimes Y\otimes I +I\otimes I\otimes Z$ where
$X\neq Y\neq Z$.

Up to now we know that any state
$\ket{\tilde{\Psi}}=\sum_{i,j,k=1}^{N}\tilde{C}_{ijk}e_{i}\otimes
e_{j}\otimes e_{k}$ can be taken by local unitary transformation
$U_1\otimes U_2 \otimes U_3$ to the state
$\ket\Psi=\sum_{i,j,k=1}^{N}C_{ijk}e_{i}\otimes e_{j}\otimes e_{k}$
where the coefficients $C_{ijk}$ fulfill (\ref{Cs}). This statement
has a deeper physical meaning. The diagonal elements of
$C^{1},C^{2},C^{3}$ constitute probabilities to obtain basis vectors
$\{e_i\}$ in some local measurements performed on state $\ket\Psi$.
The conditions (\ref{Cs}) say that any state can be transformed by
local unitary transformation to the state which is determined by
these local measurements. It is natural to ask now how to find such
a unitary local transformation.

Let us consider arbitrary state $\ket{\tilde{\Psi}}$. The action of
$U\otimes V \otimes W$ gives:
\begin{equation}\label{actPsi}
U\otimes V \otimes W
\ket{\tilde{\Psi}}=\sum_{i,j,k=1}^{N}\tilde{C}_{ijk}Ue_{i}\otimes
Ve_{j}\otimes We_{k}=\sum_{i,j,k=1}^{N}\tilde{C}_{ijk}U_{\alpha
i}V_{\beta j}W_{\gamma k}e_{\alpha}\otimes e_{\beta}\otimes
e_{\gamma}.
\end{equation}
The matrices $\tilde{C}^{1},\tilde{C}^{2},\tilde{C}^{3}$ are generally not
diagonal but by definition they are positive hence Hermitian. This means
there are unitary operators $U,V,W$ such that
$U^\dagger\tilde{C}^{1}U,V^\dagger\tilde{C}^{2}V,W^\dagger\tilde{C}^{3}W$ are
diagonal. If we take now
\begin{equation*}
C_{ijk}=\sum_{n,l,m=1}^{N}\tilde{C}_{nlm}U^T_{i n}V^T_{j l}W^T_{k
m},
\end{equation*}
then:
\begin{eqnarray*}
(C^{1})_{nl}=\sum_{j,k=1}^{N}\bar{C}_{njk}C_{ljk}=\sum_{j,k=1}^{N}\bar{\tilde{C}}_{\alpha
\beta \gamma}\bar{U^T}_{n \alpha}\bar{V^T}_{j \beta}\bar{W^T}_{k
\gamma}\tilde{C}_{abc}U^T_{l a}V^T_{j
b}W^T_{k c}=  \\
\sum_{j,k=1}^{N}\bar{\tilde{C}}_{\alpha \beta
\gamma}\tilde{C}_{abc}{U^\dagger}_{ n \alpha}U_{a l}V_{b
j}{V^\dagger}_{j \beta}W_{ c k}{W^\dagger}_{ k
\gamma}={U^\dagger}_{n \alpha}\bar{\tilde{C}}_{\alpha \beta
\gamma}\tilde{C}_{a\beta \gamma}U_{ a l}
\\
=U^\dagger_{n \alpha}(\tilde{C}^{1})_{\alpha a}{U}_{a l}=(U^\dagger
\tilde{C}^1 U)_{nl}
\end{eqnarray*}
which is diagonal as we wanted. Similarly we show that $C^2$ and
$C^3$ are diagonal as well. Now to compute the dimension of the
fiber over $\mu(\ket\Psi)$ we need to find the dimension of
submanifold of states which are sent to $\mu(\ket\Psi)$. First we
look at the coadjoint orbit through $\mu(\ket\Psi)$. As we know
$\mu(\ket\Psi)$ can be represented by an element of $X\otimes
I\otimes I+I\otimes Y\otimes I +I\otimes I\otimes Z\in\mathfrak{t}$.
Using similar reasoning as in the case of two particles we obtain:
\begin{eqnarray}\label{XYZ}
X=\mathrm{diag}(-\frac{1}{N}+p_{11}^2,-\frac{1}{N}+p_{12}^2,\ldots,
-\frac{1}{N}+p_{1N}^2), \nonumber \\
Y=\mathrm{diag}(-\frac{1}{N}+p_{21}^2,-\frac{1}{N}+p_{22}^2,\ldots,
-\frac{1}{N}+p_{2N}^2), \\
Z=\mathrm{diag}(-\frac{1}{N}+p_{31}^2,-\frac{1}{N}+p_{32}^2,\ldots,
-\frac{1}{N}+p_{3N}^2),\nonumber
\end{eqnarray}
where $\{p_{11}^2,p_{12}^2,\ldots,p_{1N}^2)\}$,
$\{p_{21}^2,p_{22}^2,\ldots,p_{2N}^2)\}$ and
$\{p_{31}^2,p_{32}^2,\ldots,p_{3N}^2)\}$ constitute the spectra of
$C^1$, $C^2$ and $C^3$, respectively. The dimension of this orbit
can be easily computed knowing that $Stab(\mu(\ket\Psi))$ consists
of matrices $U\otimes V\otimes W$ and
\begin{equation}\label{stabCC}
U=\left(
  \begin{array}{cccc}
    u_{1,0} & &  &  \\
     & u_{1,1} &  &  \\
     &  & \ddots &  \\
     &  &  & u_{1,K_1} \\
  \end{array}
\right),V=\left(
          \begin{array}{cccc}
            v_{2,0} &  &  &  \\
             & u_{2,1} &  &  \\
             &  & \ddots & \\
             &  &  & u_{2,K_2} \\
          \end{array}
        \right),
       W=\left(
          \begin{array}{cccc}
            w_{3,0} &  &  &  \\
             & w_{3,1} &  &  \\
             &  & \ddots & \\
             &  &  & w_{3,K_3} \\
          \end{array}
        \right),
\end{equation}
where $K_i$ is the number of eigenspaces of $C^i$ corresponding to
diffrent eigenvalues, $m_{i,n}$ are their dimensions and $u_{i,n}\in
U(m_{i,n})$.  Stabilizer of this orbit has dimension:
\begin{equation}\label{dimmu1}
\mathrm{dim}(Stab(\mu(\ket\Psi)))=\sum_{n=0}^{K_1}m_{1,n}^2+
\sum_{n=0}^{K_2}m_{2,n}^2+\sum_{n=0}^{K_3}m_{3,n}^2-3.
\end{equation}
Hence,
\begin{equation}\label{dimmu}
\mathrm{dim}(\mu(\mathcal{O}))=(3N^2-3)-(\sum_{n=0}^{K_1}m_{1,n}^2+
\sum_{n=0}^{K_2}m_{2,n}^2+\sum_{n=0}^{K_3}m_{3,n}^2-3).
\end{equation}
The dimension of fiber can be computed as:
\begin{equation}\label{dimmu2}
\mathrm{dim}(D(\ket\Psi))=\mathrm{dim}(Stab(\mu(\ket\Psi)))-\mathrm{dim}(Stab(\ket\Psi)).
\end{equation}
Notice that if $C^1$, $C^2$, $C^3$ have nontrivial kernels then in
decomposition of $\Psi$ there are no elements $e_i\otimes e_j\otimes
e_k$ where $e_i\in\mathrm{Ker}(C^1)$ or $e_j\in\mathrm{Ker}(C^2)$ or
$e_j\in\mathrm{Ker}(C^3)$. This means that acting on $\ket\Psi$ by
unitary operators from $Stab(\mu(\ket\Psi))$ which can be restricted
to the kernels of $C^1$, $C^2$, $C^3$ we do not change the state
$\ket\Psi$. We find thus an upper bound for the dimension of the
degeneracy space as
\begin{equation}\label{dimmu3}
\sum_{n=1}^{K_1}m_{1,n}^2+\sum_{n=1}^{K_2}m_{2,n}^2+\sum_{n=1}^{K_3}m_{3,n}^2-3.
\end{equation}
The dimension of a fiber is at least
\begin{equation}\label{lowerbound}
\max\{\sum_{n=1}^{K_1}m_{1,n}^2,\sum_{n=1}^{K_2}m_{2,n}^2,\sum_{n=1}^{K_3}m_{3,n}^2\}-1.
\end{equation}
Indeed, the conditions (\ref{Cs}) allow us to write the state
$\ket\Psi$ as
\begin{equation}\label{Cik}
\ket\Psi=\sum_{i}^{N}p_{1i}e_{i}\otimes v_i,
\end{equation}
where
\begin{equation}\label{Cik1}
v_i=\sum _{jk}\frac{1}{p_{1i}}C_{ijk}e_{j}\otimes e_{k}, \quad
i=1,\ldots,N
\end{equation}
constitute a set of orthonormal vectors. We can treat (\ref{Cik}) as a
bipartite decomposition of $\Psi$ in the orthonormal bases $\{e_i\}$ and
$\{v_i\}$. In these bases $\Psi$ is thus represented by the matrix
$\check{C}$,
\begin{gather*}
\check{C}=\left(\begin{array}{cccc}
0 I_{m_{10}}\\
 & \check{p}_{1}I_{m_{1,1}}\\
 &  & \ddots\\
 &  &  & \check{p}_{K_1}I_{m_{1,K_{1}}}\end{array}\right),
\end{gather*}
where $\check{p}_{i}$ are different eigenvalues $p_{1i}$.
Application of $U\otimes I\otimes I \in
\mathrm{Stab}(\mu(\ket{\Psi}))$ yields:
\begin{gather}\label{chCprime}
\check{C}^\prime=\left(\begin{array}{cccc}
0 u_{10}\\
 & \check{p}_{1}u_{11}\\
 &  & \ddots\\
 &  &  & \check{p}_{K_1}u_{1,K_{1}}\end{array}\right),
\end{gather}
Clearly, the matrices $\check{C}^\prime$ of the above form constitute a
manifold of dimension $\sum_{n=1}^{K_1}m_{1,n}^2-1$. In the case of two
particles this is the whole fiber because acting with $U\otimes I$ and
$I\otimes V$ we get exactly the same manifold. For multipartite systems, like
the three particle case we consider, we have to take into account that acting
with $I\otimes V\otimes I$ and $I\otimes I\otimes W$ may produce manifolds of
larger dimensionalities which leads thus to the estimate (\ref{lowerbound}).
%\begin{equation}
%max\{\sum_{n=1}^{K_1}m_{1,n}^2,\sum_{n=1}^{K_2}m_{2,n}^2,\sum_{n=1}^{K_3}m_{3,n}^2\}-1
%\end{equation}
Summing up we have
\begin{equation}\label{est}
\max\{\sum_{n=1}^{K_1}m_{1,n}^2,\sum_{n=1}^{K_2}m_{2,n}^2,\sum_{n=1}^{K_3}m_{3,n}^2\}-1\leq
D(\ket{\Psi})\leq
\sum_{n=1}^{K_1}m_{1,n}^2+\sum_{n=1}^{K_2}m_{2,n}^2+\sum_{n=1}^{K_3}m_{3,n}^2-3.
\end{equation}
Thus an orbit is symplectic if and only if
\begin{eqnarray*}\label{cond}
\sum_{n=1}^{K_1}m_{1,n}^2=1,\quad
\sum_{n=1}^{K_2}m_{2,n}^2=1,\quad\sum_{n=1}^{K_3}m_{3,n}^2=1.
\end{eqnarray*}
But this means that the state $\ket{\Psi}$ is separable because it
reduces to one of the states $e_i\otimes e_j \otimes e_k$. If the
state is separable than of course by local operations we can
transform it to the state $e_1\otimes e_1 \otimes e_1$ and then
(\ref{cond}) is fulfilled. In this way we found an easy way to check
if a state is separable and showed that it is equivalent to the fact
that associated orbit is symplectic. We also have an estimate for
dimensions of degeneracy spaces for entangled states. A
generalization to cases of more than three particles is
straightforward. So we have following
\begin{theo}
In the case of $M$ identical but distinguishable particles there is only one
symplectic orbit in the projective space
$\mathbb{P}(\bigotimes_{n=1}^M\mathbb{C}^N)$. This orbit contains all
separable states and is K\"ahler. Orbits through entangled states are not
symplectic.
\end{theo}
Knowing this and making use of Corollary \ref{collo} we arrive with
\begin{fact}
In the case of $M$ identical but distinguishable particles the dimension od
degeneracy space $D(\ket{\Psi})$ gives well defined entanglement measure.
Fora any state $\ket{\Psi}$ the estimate for this measure is given by formula
analogous to (\ref{est}).
\end{fact}

\section{Indistinguishable particles} \label{sec:bosons}

The Kostant-Sternberg theorem can be directly applied also to
indistinguishable particles, i.e.\ bosons and fermions. For $M$ bosons the
relevant group is $K=SU(N)$ represented in
$V=\mathrm{Sym}^M\left(\mathbb{C}^N\right)$. As above we want to check which
orbits of $K$-action are symplectic in the projective space $\mathbb{P}(V)$.
The best way to understand the problem is to do some nontrivial example and
then generalize the obtained result. To this end let us consider the simplest
case of $M=2$ and $N=3$, i.e. the representation of $SU(3)$ in
$\mathrm{Sym}^2\left(\mathbb{C}^3\right)$. First we notice that the
representation of $SU(N)$ in $\mathrm{Sym}^2\left(\mathbb{C}^N\right)$ is
irreducible \cite{fulton91}. From the Kostant-Sternberg theorem it follows
that it is enough to investigate structure of the
$\mathfrak{sl}(3,\mathbb{C})$ representation on
$\mathrm{Sym}^2\left(\mathbb{C}^N\right)$. The
$\mathfrak{g}=\mathfrak{sl}(3,\mathbb{C})$ algebra is eight-dimensional and
can be decomposed as $\mathfrak{g}= \mathfrak{n}_-\oplus \mathfrak{h}\oplus
\mathfrak{n}_+$, where $\mathfrak{h}$ is the Cartan subalgebra consisting of
traceless diagonal matrixes and $\mathfrak{n}_+=\mathrm{Span}(
E_{12},E_{13},E_{23})$, $\mathfrak{n}_-=\mathrm{Span}(E_{21},E_{31},E_{32})$.
We define three linear functionals $L_i:\mathfrak{h}\rightarrow\mathbb{C}$,
\begin{equation}\label{Lweight}
L_i(diag(a_1,a_2,a_3))=a_i, \quad i\in \{1,2,3\}
\end{equation}

Let us choose a basis $B$ in $\mathrm{Sym}^2\left(\mathbb{C}^3\right)$,
$B=\{e_1\otimes e_1,\ e_2\otimes e_2,\ e_3\otimes e_3, e_1\otimes
e_2+e_2\otimes e_1,\ e_1\otimes e_3+e_3\otimes e_1,\ e_2\otimes
e_3+e_3\otimes e_2\}$ where $e_i\in \mathbb{C}^3$ are the standard basis
vectors. The action of the Lie algebra $\mathfrak{g}$ on
$\mathrm{Sym}^2\left(\mathbb{C}^3\right)$ is a usual action of the tensor
product of representations. Construction of the representation of
$\mathfrak{sl}(3,\mathbb{C})$ on $\mathrm{Sym}^2\left(\mathbb{C}^3\right)$ is
straightforward, we take the vector $e_1\otimes e_1$ which is the highest
weight vector (it is an eigenvector of all elements in $\mathfrak{h}$ and it
is annihilated by $\mathfrak{n}_+$ ), and we act on it with operators from
$\mathfrak{n}_-$. As a result we obtain a decomposition of
$V=\mathrm{Sym}^2\left(\mathbb{C}^3\right)$ into the direct sum $V=\bigoplus
V_\lambda$ where the one-dimensional weight spaces $V_\lambda$ are spanned by
the basis vectors of $B$. The weights $\lambda\in \mathfrak{h}^\ast$ can be
now calculated as \footnote{Remember that we represent $\mathfrak{k}$ in the
symmetric tensor product, hence $H(e_i\otimes e_i)$ has the meaning of
$(H\otimes I+I\otimes H)(e_i\otimes e_i)=He_i\otimes e_i+e_i\otimes He_i$
etc.}
\begin{eqnarray}
H(e_i\otimes e_i)=2L_i(H)e_i\otimes e_i\qquad i=1,2,3,  \nonumber \\
H(e_i\otimes e_j+e_j\otimes e_i)=(L_i+L_j)(H)\left(e_i\otimes
e_j+e_j\otimes e_i\right)
\end{eqnarray}
We know that only orbits passing through weight vectors might be symplectic.
We have the following $\mathfrak{sl}(2,\mathbb{C})$ triples in
$\mathfrak{sl}(3,\mathbb{C})$: $(E_{ij},E_{ji},H_{ij}=[E_{ij},E_{ji}])$. The
orbit through a weight vector $v$ with a weight $\lambda$ is symplectic if
and only if for every operator from $\mathfrak{n}_+$ the following
implication is true: $\lambda(H_{ij})=0 \Rightarrow E_{ij}(v)=0=E_{ji}(v)$.
There are two cases to consider,
\begin{itemize}
\item Vectors of the form $e_i\otimes e_i$, $i=1,2,3$. The weight of the
    $e_i\otimes e_i$ state is $2L_i$ so $2L_i(H_{kj})=0$ only when $k\neq
    i$ and $j\neq i$. In this case $E_{kj}(e_i\otimes e_i)=0$ because to
    give nonzero result matrix $E_{kj}$ must have one in the $i$-th
    column. The corresponding orbit is thus symplectic. Obviously, all
    these vectors lie on the orbit through the highest weight vector
    $e_1\otimes e_1$
\item Vectors of the form $e_i\otimes e_j+e_j\otimes e_i$ where $i\neq
    j$.The weight of this vector is $L_i+L_j$ so $(L_i+L_j)(H_{kl})=0$
    only if $k=i$ and $j=l$. In this case $E_{ij}(e_i\otimes
    e_j+e_j\otimes e_i)\neq0$ because $E_{ij}e_j\neq0$. In conclusion the
    orbit through $e_i\otimes e_j+e_j\otimes e_i$ is not symplectic.
\end{itemize}

Let us now return to the problem mentioned in Section~\ref{subsec:ind-def}.
If we define nonentangled bosonic states as antisymmetrizations of simple
tensors (or, more precisely, as corresponding points in the projective space)
then we clearly have two, inequivalent from the geometric point of view,
types of nonentanglement. Non-entangled states of two different types are not
connected by local unitary transformations which is in contrast to the
familiar situation of distinguishable particles and intuitions build upon the
fact that all separable states of distinguishable particles can be obtained
from a single one by local transformations. Although this is obviously
acceptable, it remains an open problem what is a physical meaning of two
different types of nonentanglement. If, instead, we adopt the second
definition identifying nonentangled bosonic states as points in the
projective space corresponding to tensor products of the same vector we
encounter the same situation as in the case of distinguishable particles -
the nonentangled states form a unique symplectic orbit, and the degeneracy of
the symplectic form can be used as a measure of entanglement for entangled
states.

The case of fermions does not lead to any ambiguities of the above type.
Calculations similar to those made for bosons lead to a conclusion that
nonentangled states form the unique symplectic orbit. Indeed, let us consider
as an example $K=SU(N)$ and $V=\bigwedge^2\mathbb{C}^N$ corresponding to two
fermions of spin $(N-1)/2$ with the single-particle space
$\mathcal{H}_1=\mathbb{C}^N$. In terms of the previously introduced standard
bases $e_i$ and $E_{ij}$ adapted to $N$ dimensions $V$ is spanned by
$e_{kl}=e_k\otimes e_l-e_l\otimes e_k$, with $k<l$ the highest weight vector
is $e_{12}$ and
\begin{equation}\label{staction}
E_{ij}\,e_{kl}=\delta_{jl}e_{ki}+\delta_{jk}e_{il}
\end{equation}
where we denote $e_{kl}=-e_{lk}$ for $k>l$. Acting by $E_{ij}$ with $i>j$ on
$e_{12}$ we obtain remaining weight vectors, which according to
(\ref{staction}) are all of the form $e_{kl}$ (in fact with $l=1$ or $2$),
with weights $L_i+L_j$ (we extended in an obvious way the definition
(\ref{Lweight}) to $N$ dimensions). As remarked $(L_i+L_j)(H_{kl})=0$ implies
$k=i$, $j=l$ but then $E_{ij}e_{kl}=0=E_{ji}e_{kl}$.

\section{Summary and outlook}

We presented an geometric description of the set of pure states of composite
quantum systems in terms of natural symplectic structure in the space of
states. Nonentangled states form a unique symplectic orbit through the
highest weight vector of the appropriate representation of the group of local
transformations whereas entangled states are characterized by the degeneracy
of the symplectic form. The degeneracy can be thus used as a kind of
geometric measure of entanglement. We were able to calculate the degeneracy
in many relevant cases and give some estimates for the most general system of
arbitrary number of constituents with an arbitrary dimension of the single
particle space. Let us remark that there exists a useful characterization of
the highest weight vector orbits which allows to generalize and estimate
effectively some other entanglement measures \cite{kkk10}.

An obvious question is whether a method can be adapted to the case of mixed
states. This problem, as well as applications of the obtained results to
identifying, so called, locally unitary equivalent multiparticle states
\cite{kraus10} and finding "canonical" forms of them we postpone to
forthcoming publications.

\section{Acknowledgments}
The support by SFB/TR12 'Symmetries and Universality in Mesoscopic
Systems' program of the Deutsche Forschungsgemeischaft and Polish
MNiSW grant no.\ DFG-SFB/38/2007 is gratefully acknowledged.

\section{Appendix}
\subsection{Symplectic structure on coadjoint orbits}
Let $K$ be a semisimple compact Lie group, $\mathfrak{k}$ its Lie algebra,
and $\mathfrak{k}^\ast$ the dual space to $\mathfrak{k}$. The coadjoint
action of $K$ on $\mathfrak{k}^\ast$ is given by
\begin{gather}\label{cadjoint}
\mathrm{Ad}^\ast_g:\mathfrak{k}^\ast \rightarrow\mathfrak{k}^\ast\\\nonumber
\langle\mathrm{Ad}^\ast_g\alpha,Y\rangle=\langle\alpha,\mathrm{Ad}_{g^{-1}}
Y\rangle=\langle\alpha,g^{-1}Yg\rangle, \quad g\in K, \quad Y\in\mathfrak{k},
\quad \alpha\in\mathfrak{k}^\ast,
\end{gather}
where $\mathrm{Ad}$ is the adjoint action of $K$. It can be easily checked
that (\ref{cadjoint}) is well defined. The coadjoint orbit
$\mathcal{O}_\alpha$ passing through $\alpha\in \mathfrak{k}^\ast$ is the
orbit of the coadjoint action of $K$ on $\alpha$
\begin{equation}\label{coadorbit}
\mathcal{O}_\alpha=\{\mathrm{Ad}^\ast_g\alpha:g\in K\}
\end{equation}
Our goal now is to define a $K$-invariant symplectic form $\omega$ on
$\mathcal{O}_\alpha$. Such a form acts on tangent vectors so it is reasonable
to first look at their structure. Since $\mathfrak{k}^\ast$ is a vector space
its tangent space at any point is again $\mathfrak{k}^\ast$. For any $X\in
\mathfrak{k}$ let $\tilde{X}\in T_\alpha\mathcal{O}_\alpha$ be a vector
tangent to the curve $t\mapsto \mathrm{Ad}^\ast_{exp(tX)}\alpha$. We have
then,
\begin{gather}\label{tangentvector}
\langle\tilde{X},Y\rangle=\langle\frac{d}{dt}\bigg|_{t=0}\mathrm{Ad}^\ast_{exp(tX)}\alpha,Y\rangle=\frac{d}{dt}\bigg|_{t=0}\langle\alpha,\mathrm{Ad}_{exp(-tX)}Y\rangle=\langle\alpha,[Y,X]\rangle,
\end{gather}
where $Y\in\mathfrak{k}$. Thus $\tilde{X}$ is an element of
$\mathfrak{k}^\ast$ given by
\begin{equation}\label{tildeX}
\tilde{X}=\langle\alpha,[\,\cdot\,,X]\rangle.
\end{equation}
It is now interesting to ask how this tangent vector transform when pushed by
an element $g\in K$, i.e.\ to consider the vector $g\tilde{X}\in
T_{\mathrm{Ad}^\ast_g\alpha}\mathcal{O}_\alpha$ tangent to the curve
$t\mapsto \mathrm{Ad}^\ast_g\mathrm{Ad}^\ast_{exp(tX)}\alpha$. We have
\begin{gather}\label{pushtangent}
\langle
g\tilde{X},Y\rangle=\langle\frac{d}{dt}\bigg|_{t=0}\mathrm{Ad}^\ast_g\mathrm{Ad}^\ast_{exp(tX)}\alpha,Y\rangle=\frac{d}{dt}\bigg|_{t=0}\langle\alpha,\mathrm{Ad}_{exp(-tX)}\mathrm{Ad}_{g^{-1}}Y\rangle=\\\nonumber=\langle\alpha,[\mathrm{Ad}_{g^{-1}}Y,X]\rangle
=\langle\alpha,[\mathrm{Ad}_{g^{-1}}Y,\mathrm{Ad}_{g^{-1}}\mathrm{Ad}_{g}X]=\\\nonumber
=\langle\alpha,\mathrm{Ad}_{g^{-1}}[Y,\mathrm{Ad}_{g}X]\rangle=\langle\mathrm{Ad}^\ast_g\alpha,[Y,\mathrm{Ad}_{g}X]\rangle.
\end{gather}
Hence $g\tilde{X}$ is an element of $\mathfrak{k}^\ast$ given by
\begin{equation}\label{tildegX}
g\tilde{X}=\langle\mathrm{Ad}^\ast_g\alpha,[\,\cdot\,,\mathrm{Ad}_gX]\rangle.
\end{equation}
We may now define our symplectic form at any $\beta\in\mathcal{O}_\alpha$ as
\begin{equation}\label{symformcoa}
\omega_\beta(\tilde{X},\tilde{Y})=\langle\beta,[X,Y]\rangle.
\end{equation}
The form is non-degenerate on $\mathcal{O}_\alpha$ because
\begin{gather}\label{ndegenerate}
\forall Y\in\mathfrak{k}\quad\langle\beta,[X,Y]\rangle=0\Leftrightarrow
\forall Y \in\mathfrak{k} \quad
\langle\tilde{X},Y\rangle=0\Leftrightarrow\tilde{X}=0.
\end{gather}
It is also $K$-invariant because
\begin{gather}\label{ginvariant}
\omega_{\mathrm{Ad}^\ast_g\beta}(g\tilde{X},g\tilde{Y})=\langle\mathrm{Ad}^\ast_g\beta,[Ad_gX,Ad_gY]\rangle=\langle\beta,Ad_{g^{-1}}Ad_g[X,Y]\rangle=\\\nonumber
=\langle\beta,[X,Y]\rangle=\omega_\beta(\tilde{X},\tilde{Y}).
\end{gather}
Thus we need only to check if $\omega$ is closed. But we know that for any
differential $2$-form the following is true
\begin{gather}\label{domega}
d\omega(\tilde{X},\tilde{Y},\tilde{Z})=\tilde{X}\omega(\tilde{Y},\tilde{Z})+\tilde{Y}\omega(\tilde{Z},\tilde{X})+\tilde{Z}\omega(\tilde{X},\tilde{Y})+\\\nonumber
-(\omega([\tilde{X},\tilde{Y}],\tilde{Z})+\omega([\tilde{Y},\tilde{Z}],\tilde{X})+\omega([\tilde{Z},\tilde{X}],\tilde{Y})).
\end{gather}
Taking
$\tilde{X}=\langle\mathrm{Ad}^\ast_g\beta,[\,\cdot\,,\,\mathrm{Ad}(g)X\,]\rangle$
and similarly for $\tilde{Y}$ and $\tilde{Z}$ we see that that first three
terms in (\ref{domega}) vanish because $\omega$ is $K$-invariant. The sum of
next three terms is also zero due to the Jacobi identity. This way we arrive
at a well defined symplectic form on $\mathcal{O}_\alpha$.
\subsection{K\"{a}hler structure}
Let us start with definition of K\"{a}hler manifold.
\begin{defi}
 Let $M$ be a complex
manifold $\mathrm{dim}_{\mathbb{C}}M=n$ and let $\omega$ be a symplectic form
on $M$ treated as $2n$-dimensional real manifold. Then $M$ is called a
K\"{a}hler manifold if at every $p\in X$ the complex structure $i$ on $T_pX$
(multiplication by imaginary unit) and the antisymmetric form $\omega_p$ has
the following property:
\begin{equation}\label{kaehler}
    \omega_p(iv,iw)=\omega(v,w),
\end{equation}
that is $i\in Sp(T_pM)$ (the symplectic group of $T_pM$).
\end{defi}
Assume $M$ is a K\"{a}hler manifold. Then we can define a symmetric
nondegenerate form $b$ on $M$
\begin{equation}\label{bform}
b(v,w)=\omega(v,iw).
\end{equation}
Indeed we have
\begin{equation}\label{bformcheck}
b(v,w)=\omega(v,iw)=\omega(iv,i^2w)=-\omega(iv,w)=\omega(w,iv)=b(w,v).
\end{equation}
The form $b$ is non-degenerate because
\begin{equation}\label{bnondegen}
\forall w\in T_pM\quad b(v,w)=0\Leftrightarrow \forall w\in
T_pM\quad \omega(w,iv)=0\Leftrightarrow v=0.
\end{equation}
It is also $i$-invariant
\begin{equation}\label{iinvariant}
b(iv,iw)=\omega(iv,i^2w)=\omega(v,iw)=b(v,w).
\end{equation}
We have now the following definition.
\begin{defi}
A K\"{a}hler structure on $M$ is called positive if and only if the
corresponding symmetric form $b$ is positive.
\end{defi}
It is straightforward to check that having such an $i$-invariant
non-degenerate symmetric form $b$ on $M$ we can a define non-degenerate
$i$-invariant antisymmetric $2$-form by $\omega(v,w)=b(iv,w)$. We will need
only one more theorem (see \cite{guillemin84} for a detailed proof).
\begin{theo}\label{apptheo}
Let $M$ be a positive K\"{a}hler manifold. Then any complex submanifold
$N\subset M$ is also a K\"{a}hler manifold.
\end{theo}
The assumption that $M$ is a positive K\"{a}hler manifold (not just a
K\"{a}hler manifold) is very important due to the fact that restriction of
the symmetric form $b$ to $N$ is then non-degenerate and $i$-invariant. Now
using $b|_N$ we can define a non-degenerate $i$-invariant antisymmetric
$2$-form on $N$ which is a restriction of $\omega$ defined on the whole $M$,
and hence is symplectic.

\subsubsection{K\"{a}hler structure on $\mathbb{P}(V)$}
Consider a complex vector space $V$, $\mathrm{dim}_\mathbb{C}V=n$ with a
Hermitian scalar product $(\cdot|\cdot)$. The complex projective space
$\mathbb{P}(V)$ of $V$ is defined
\begin{equation}\label{cpspace}
\mathbb{P}(V)=V/\sim,
\end{equation}
where the equivalence $\sim$ is defined by
\begin{equation}\label{sim}
v\sim w\Leftrightarrow w=\alpha v\quad\alpha\in\mathbb{C}^\ast,
\end{equation}
where $\mathbb{C}^\ast=\mathbb{C}\setminus\{0\}$. The standard way
to realize $\mathbb{P}(V)$ is by two steps
\begin{equation}\label{twosteps}
V\stackrel{a}{\longrightarrow}S(V)\stackrel{b}{\longrightarrow}
\mathbb{P}(V),
\end{equation}
where the step $a$ is a quotient by the dilation $v\sim \alpha v$ for $a\in
\mathbb{R}^\ast$ and the step $b$ is a quotient by the rotations $v\sim
e^{i\phi}v$. The result of the quotient $a$ is the real sphere $S(V)=\{v\in
v:\|v\|=1\}$. The quotient $b$ gives $S(V)/S^1=S^{2n-1}/S^1$, where $S^1$
represents the group of rotations. It is well known fact that the complex
projective space $\mathbb{P}(V)$ is a complex manifold and
$\mathrm{dim}_\mathbb{C}\mathbb{P}(V)=n-1$. Let
$\pi:V\setminus\{0\}\rightarrow \mathbb{P}(V)$ be the projection defined by
equivalence $\sim$. The tangent space to $\mathbb{P}(V)$ at $z=\pi(v)$ is
$T_z\mathbb{P}(V)=\pi_v(T_vV)$. It is good question to ask what is
$\pi_v(\xi)$, where $\xi\in T_vV$. Consider the curve $t\mapsto v(t)\in V$,
$v(0)=v$. Then
\begin{equation}\label{xivect}
\xi=\frac{d}{dt}\bigg|_{t=0}v(t).
\end{equation}
We first apply the map $a$ to $v(t)$. As a result we get a curve $t\mapsto
\frac{v(t)}{\|v(t)\|}\in S(V)$. Applying the map $b$ amounts to getting rid
of the rotation $e^{i\phi}v$, hence finally the curve in $\mathbb{P}(v)$ is
given as
\begin{equation}\label{PVcurve}
t\mapsto
\frac{v(t)}{\|v(t)\|}-v\big(\frac{v}{\|v\|}\big|\frac{v(t)}{\|v(t)\|}\big).
\end{equation}
The tangent vector to this curve is thus
\begin{gather}\label{xivectSV}
\frac{d}{dt}\bigg|_{t=0}\frac{v(t)}{\|v(t)\|}-\frac{d}{dt}\bigg|_{t=0}v\big(\frac{v}{\|v\|}\big|\frac{v(t)}{\|v(t)\|}\big)=\frac{\xi}{\|v\|}-\frac{v}{\|v\|}\big(\frac{v}{\|v\|}\big|\frac{\xi}{\|v\|}\big).
\end{gather}
Hence for any vector $\xi\in T_vV$ we obtain the corresponding vector
$\pi_v(\xi)\in T_{\pi(v)}\mathbb{P}(V)$ as the orthogonal complement of
$\frac{\xi}{\|v\|}$ to the subspace $\mathbb{C}v$ in the Hermitian scalar
product $(\cdot|\cdot)$. Let us introduce a Hermitian scalar product on
$\mathbb{P}(v)$ by
\begin{gather}\label{hproduct}
h(\pi_v(\xi),\pi_v(\eta))=\bigg(\frac{1}{\|v\|}\frac{\xi(v|v)-(v|\xi)v}{(v|v)}\bigg|\frac{1}{\|v\|}\frac{\eta(v|v)-(v|\eta)v}{(v|v)}\bigg),
\end{gather}
which, after some calculations, reduces to
\begin{equation}\label{hproductbis}
h(\pi_v(\xi),\pi_v(\eta))=\frac{(\xi|\eta)(v|v)-(\xi|v)(v|\eta)}{(v|v)^2}.
\end{equation}
Of course, $h$ is a well defined, that is non-degenerate, positive Hermitian
form on $\mathbb{P}V$. Indeed, for any $z=\pi(v)$ we have
$T_{z}\mathbb{P}(V)=(\mathbb{C}v)^\perp$. Knowing this we can introduce a
non-degenerate antisymmetric $2$-form $\omega$ on $\mathbb{P}(V)$ as the
imaginary part of $h(\cdot,\cdot)$
\begin{equation}\label{om}
\omega(\pi_v(\xi),\pi_v(\eta))=-\mathrm{Im}h(\pi_v(\xi),\pi_v(\eta)).
\end{equation}
It is straightforward to check that $\omega$ is not only $i$-invariant but
also $U(V)$-invariant. To check that it is also closed notice that $U(V)$ is
acting transitively on $V$ which means the vectors $Av$, where $A\in u(V)$
span $T_vV$. Hence,
\begin{equation}\label{rest}
\omega(\pi_v(Av),\pi_v(Bv))=-\mathrm{Im}\frac{(Av|Bv)(v|v)-(Av|v)(v|Bv)}{(v|v)^2},\quad
A,B\in u(V).
\end{equation}
But $(Av|v)(v|Bv)$ is real since $(Av|v)$ and $(v|Bv)$ are imaginary (this is
because $A^\dagger=-A$ for all $A\in u(V)$). Hence,
\begin{equation}\label{ome}
\omega(\pi_v(Av),\pi_v(Bv))=-\mathrm{Im}\frac{(Av|Bv)}{(v|v)}=\frac{i([A,B]v|v)}{2(v|v)},\quad
A,B\in u(V).
\end{equation}
Now making use of Equation~(\ref{domega}) for vector fields
$\pi_{Uv}(UAU^\ast Uv)$ and similarly for $B$ and $C$ we find, by the same
argument as for coadjoint orbits, that $d\omega=0$. This means
$\mathbb{P}(V)$ is a K\"{a}hler manifold. It is also a positive K\"{a}hler
manifold because the corresponding symmetric form $b$ given by
\begin{equation}\label{PVbform}
b(\pi_v(\xi),\pi_v(\eta))=-\mathrm{Re}h(\pi_v(\xi),\pi_v(\eta)),
\end{equation}
is clearly positive.

%
%\bibliographystyle{unsrt}
%\bibliography{symp}

\end{document}